\documentclass[fleqn,usenatbib]{mnras}

\usepackage{amsfonts}

\usepackage[T1]{fontenc}
\usepackage{ae,aecompl}
\usepackage[position=top]{subfig}
\usepackage{float}
\usepackage{comment}

\usepackage{grffile}
\usepackage{etoolbox}
\makeatletter 
  \patchcmd{\NAT@citex}
    {\@citea\NAT@hyper@{%
      \NAT@nmfmt{\NAT@nm}%
      \hyper@natlinkbreak{\NAT@aysep\NAT@spacechar}{\@citeb\@extra@b@citeb}%
      \NAT@date}}
    {\@citea\NAT@nmfmt{\NAT@nm}%
    \NAT@aysep\NAT@spacechar\NAT@hyper@{\NAT@date}}{}{}

  \patchcmd{\NAT@citex}
    {\@citea\NAT@hyper@{%
      \NAT@nmfmt{\NAT@nm}%
      \hyper@natlinkbreak{\NAT@spacechar\NAT@@open\if*#1*\else#1\NAT@spacechar\fi}%
        {\@citeb\@extra@b@citeb}%
      \NAT@date}}
    {\@citea\NAT@nmfmt{\NAT@nm}%
    \NAT@spacechar\NAT@@open\if*#1*\else#1\NAT@spacechar\fi\NAT@hyper@{\NAT@date}}
    {}{}
\makeatother

\usepackage{etoolbox}
 \makeatother
\usepackage{graphicx}	
\usepackage{amsmath}	
\usepackage{amssymb}	
\usepackage{pifont}
\title[GWs from Pop~III remnants in NSCs]{Gravitational waves from the remnants of the first stars in nuclear star clusters}

\author[B. Liu \& V. Bromm]{Boyuan Liu\textsuperscript{\href{https://orcid.org/0000-0002-4966-7450}{\includegraphics[width=2.5mm]{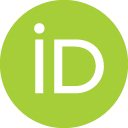}}\,}\thanks{E-mail: boyuan@utexas.edu}$^{1}$ 
and Volker Bromm$^{1}$
\\
$^{1}$Department of Astronomy, University of Texas, Austin, TX 78712, USA\\
}

\date{Accepted XXX. Received YYY; in original form ZZZ}

\pubyear{2021}

\begin{document}
\label{firstpage}
\pagerange{\pageref{firstpage}--\pageref{lastpage}}
\maketitle

\begin{abstract}
We study Population~III (Pop~III) binary remnant mergers in nuclear star clusters (NSCs) with a semi-analytical approach for early structure formation. Within this framework, we keep track of the dynamics of Pop~III binary (compact object) remnants during cosmic structure formation, and construct the population of Pop~III binary remnants that fall into NSCs by dynamical friction of field stars. The subsequent evolution within NSCs is then derived from three-body encounters and gravitational-wave (GW) emission. 
We find that 7.5\% of Pop~III binary remnants will fall into the centres ($< 3\ \rm pc$) of galaxies. About $5-50\%$ of these binaries will merge at $z>0$ in NSCs, including those with very large initial separations (up to 1~pc). The merger rate density (MRD) peaks at $z\sim 5-7$ with $\sim 0.4-10\ \rm yr^{-1}\ \rm Gpc^{-3}$, leading to a promising detection rate $\sim 170-2700\ \rm yr^{-1}$ for 3rd-generation GW detectors that can reach $z\sim 10$. 
Low-mass ($\lesssim 10^{6}\ \rm M_{\odot}$) NSCs formed at high redshifts ($z\gtrsim 4.5$) host most ($\gtrsim 90$\%) of our mergers, which mainly consist of black holes (BHs) with masses $\sim 40-85\ \rm M_{\odot}$, similar to the most massive BHs found in LIGO events. Particularly, our model can produce events like GW190521 involving BHs in the standard mass gap for pulsational pair-instability supernovae with a MRD $\sim 0.01-0.09\ \rm yr^{-1}\ Gpc^{-3}$ at $z\sim 1$, consistent with that inferred by LIGO. 
\end{abstract}



\section{Introduction}
\label{s1}
The detection of gravitational waves (GWs) from mergers of compact objects, such as black holes (BHs) and neutron stars (NSs), has opened a new observational window in astrophysics, cosmology and fundamental physics (reviewed by, e.g. \citealt{barack2019black}). These ripples of spacetime carry valuable information for a variety of astrophysical processes, such as formation and evolution of compact object binaries, BH mass function, cosmic expansion and structure/star formation (e.g. \citealt{dvorkin2016metallicity,fishbach2018does,mapelli2019properties,vitale2019measuring,perna2019constraining,safarzadeh2019measuring,farr2019future,adhikari2020binary,tang2020dependence,safarzadeh2020branching,bouffanais2021new,deluca2021bayesian,fishbach2021when,mastrogiovanni2021cosmology,wang2021black}). The GW window ideally complements the electromagnetic (EM) window with two advantages that are particularly important at high redshifts: (i) EM signals decay rapidly with (luminosity) distance ($\propto d_{L}^{-2}$) such that EM observations at high-$z$ are significantly biased towards massive/luminous systems, while the amplitudes of GWs decay slower ($\propto d_{L}^{-1}$), making it easier for GW observations to reach high-$z$. (ii) Compact object binaries formed in the early Universe can merge at later times such that imprints of high-$z$ processes can be inferred from GW events detected in the local Universe.

These unique features of GWs make them a promising probe to the first stars, the so-called Population~III (Pop~III) with extremely low or zero metallicity ($Z\lesssim 10^{-6}-10^{-4}\ \rm Z_{\odot}$, reviewed by e.g. \citealt{bromm2009formation,bromm2013,haemmerle2020formation}), for which direct observations in the EM window will be challenging even with the \textit{James Webb Space Telescope (JWST)} \citep{gardner2006james,anna2020tele}. 
Compared with Population~I/II (Pop~I/II) stars formed in metal-enriched gas, Pop~III stars produce compact object remnants much more efficiently because of their lack of strong mass loss and top-heavy initial mass function (IMF) resulting from the insufficient cooling of primordial gas. This makes them ideal progenitors of compact object mergers that can potentially account for a significant fraction of the GW events detected at $z\lesssim 1$ so far by LIGO \citep{3ogc}, and those to be discovered at higher redshifts. The recent detection of the special event GW190521 with unusual BH masses $85_{-14}^{+21}\ \rm M_{\odot}$ and $66_{-18}^{+17}$ \citep{abbott2020gw190521,abbott2020properties} further highlights the importance of Pop~III remnants. These BH masses are forbidden for Pop~I/II by standard pulsational pair-instability supernova (PPISN) models (e.g. \citealt{heger2003massive,belczynski2016effect,woosley2017pulsational,marchant2019pulsational}). Pop~III stars, on the other hand, can produce BHs in this mass range by retaining the hydrogen envelope before collapse and avoiding the PPISN regime due to their compactness and strongly reduced mass loss \citep{farrell2020gw190521,tanikawa2020population}. Furthermore, the observed merger rate density (MRD) for GW190521-like events can also be reproduced considering a Pop~III origin \citep{kinugawa2021formation,bl2020gw190521}.

The GW signals from Pop~III binary remnant mergers have been intensively studied for the classical binary stellar evolution (BSE) channel (e.g. 
\citealt{kinugawa2014possible,kinugawa2015detection,hartwig2016,belczynski2017likelihood,tanikawa2021merger,hijikawa2021population,kinugawa2021}). In this scenario for close binaries in isolation, interactions such as mass transfer, tidal effects, and common envelope evolution shrink the binary orbits to facilitate merging within a Hubble time. The predictions from these studies exhibit significant discrepancies due to different assumptions on the Pop~III star formation (SF) history, initial binary statistics and BSE parameters. For instance, even if the Pop~III SF history is constrained, the predicted local ($z\sim 0$) MRDs have large scatters of two orders of magnitude, such that it is unclear which fraction (none or up to 50\% ) of the LIGO events can be attributed to Pop~III remnants.

Unfortunately, it is difficult to reduce the uncertainties in the binary statistics and BSE parameters without direct observations of Pop~III stars. Besides, all these studies rely on the assumption that Pop~III binaries are similar to their Pop~I/II counterparts (with observational constraints), which can be an over-idealization, given that Pop~III SF occurs under markedly different conditions (see e.g. \citealt{chon2021transition}). Actually, recent high-resolution (radiative) hydrodynamic simulations of Pop~III SF and N-body simulations of young Pop~III star clusters \citep{susa2019merge,sugimura2020birth,liu2021binary} indicate that close binaries of Pop~III stars are likely rare because of the expansion of the system by angular momentum conservation during protostellar accretion.

The challenges for the BSE channel motivate us to explore alternative pathways of forming Pop~III binary remnant mergers that can also work even if Pop~III binaries are initially wide (with separations $\gtrsim 10\ \rm au$). Dynamical hardening (DH) in nuclear star clusters (NSCs) has been shown to be an efficient mechanism of shrinking binary orbits for Pop~I/II remnants and supermassive BHs (see e.g. \citealt{sesana2015scattering,antonini2016merging,choksi2019star,sedda2020birth,mapelli2021hierarchical}). In this scenario, massive, hard binaries of compact objects will be further hardened by three-body encounters with surrounding stars, which can drive them to mergers within a Hubble time under the high stellar densities of NSCs. In our case, Pop~III (binary) remnants need to fall into NSCs for this mechanism to work, which is a complex process involving the dynamics of Pop~III remnants during cosmic structure formation. The exploratory study by \citet{bl2020gw190521} applies this NSC-DH channel to Pop~III remnants, based on a simple model for the dynamics of Pop~III remnants in their host galaxies, finding that it is possible to explain the GW190521 event. In this work, we design a more advanced model to keep track of the motions of Pop~III remnants driven by dynamical friction (DF) from field stars, coupled with a semi-analytical framework for early structure formation based on halo merger trees. By taking into account the cosmological context for the dynamics of Pop~III remnants in their host galaxies and during galaxy mergers, we can predict when and where Pop~III remnants fall into NSCs. The subsequent evolution of Pop~III binary remnants inside NSCs is then followed by post-processing to predict their mergers. 

This paper is organised as follows. In Section~\ref{s2}, we describe our semi-analytical framework for early structure formation (Sec.~\ref{s2.1}) and models for the in-fall of Pop~III binary remnants into NSCs (Sec.~\ref{s2.2}) and their subsequent evolution (Sec.~\ref{s2.3}). In Section~\ref{s3}, we present our main results for the properties of host galaxies/NSCs at the moments of NSC in-fall (Sec.~\ref{s3.1}) and mergers of Pop~III binary remnants in NSCs, compared with previous theoretical predictions for the BSE channel and LIGO observations (Sec.~\ref{s3.2}). Finally, in Section~\ref{s4}, we summarize the key features of the NSC-DH channel for Pop~III binary remnant mergers and discuss their implications for GW astronomy in the early Universe.

\section{Methodology}
\label{s2}
In this section, we present our method of modelling Pop~III binary remnant mergers in (high-$z$) NSCs. We start with a merger-tree based framework for early structure formation that self-consistently models SF, stellar feedback, and particularly, Pop~III binary stars and the relevant binary remnants (Sec.~\ref{s2.1}). Then we keep track of the dynamics of such Pop~III binary remnants in their host galaxies, modelling the population of binaries falling into NSCs by DF from field (low-mass Pop~I/II) stars (Sec.~\ref{s2.2}). Finally, with an observation-based NSC model we follow the evolution of such binaries in NSCs (Sec.~\ref{s2.3}) via DH and (potential) ejection by three-body encounters, as well as GW emission. The binary properties involved are derived from N-body simulations of Pop~III star clusters in \citet{liu2021binary}.

\subsection{Early structure formation}
\label{s2.1}

Following earlier work \citep{hartwig2015constraining,hartwig2016}, we model early structure formation with halo merger trees generated by the \textsc{galform} code \citep{parkinson2008generating}. For simplicity, we simulate 500 trees starting from an initial redshift of $z_{i}=42$, each targeted at a Milky-Way (MW)-like halo of total mass $M_{\rm h}=1.26\times 10^{12}\ \rm M_{\odot}$ at $z_{\rm vir}\simeq 2.5$. The mass resolution of the trees is $\Delta M_{\rm h}=5\times 10^{5}\ \rm M_{\odot}$, and the redshift resolution is $\Delta z\simeq 0.15$, corresponding to timesteps of $\sim 0.3-180\ \rm Myr$. {Since such MW-like haloes with $M_{\rm h}\sim 10^{12}\ \rm M_{\odot}$, corresponding to $\sim 1-2\sigma$ peaks at $z\lesssim 2.5$, are massive enough to capture small-scale perturbations that collapse at high-$z$, we expect their average assembly histories to be cosmologically representative before turn-over ($z\gtrsim 5$), when most Pop~III stellar remnants are formed. Our results for the low-redshift regime ($z\lesssim 5$) are derived from simple assumptions and extrapolation (see Sec.~\ref{s2.2}), which may be incomplete. As the goal of this study is to demonstrate the NSC-DH channel for Pop~III remnant mergers, 
we defer more accurate modelling at low-$z$ with complete halo populations to future work.} 

The merger trees are coupled with a customized model for primordial SF that self-consistently takes into account metal enrichment and Lyman-Werner (LW) feedback (for details, see \citealt{hartwig2015constraining}, HT15 henceforth). We have updated this model to achieve better agreement with observations \citep{campisi2011population,madau2014araa,aghanim2020planck}, and implemented recipes for Pop~I/II SF, important for the dynamics of Pop~III remnants in their host galaxies. Below we describe our main refinements to the HT15 model.

Same as HT15, we use the cosmological parameters from \textit{Planck} 2013 results \citep{ade2014planck} for a flat $\Lambda$CDM universe: $H_{0}=67.77\ \rm km\ s^{-1}\ Mpc^{-1}$, $\Omega_{\rm m}=0.309$, $\Omega_{\rm b}=0.048$, $n_{\rm s}=0.9611$ and $\sigma_{8}=0.8288$. The dark matter power spectrum is calculated with the \textsc{camb} code \citep{lewis2000efficient} in the wave number ($k$) range $k/h\sim 10^{-6}-10^{6}\ \mathrm{Mpc^{-1}}$, where $h\equiv H_{0}/(100\ \rm km\ s^{-1}\ Mpc^{-1})$

\subsubsection{Pop~III star formation}
\label{s2.1.1}

The original Pop~III SF model in HT15 does not consider the cooling/collapse timescale of the star-forming cloud in the host halo, and therefore, significantly overproduces Pop~III stars at very high redshifts ($z\sim 15-35$) by instantaneous SF. This leads to early reionization ($z\gtrsim 8$) with a high Thomson optical depth $\tau\gtrsim 0.09$, inconsistent with the recent measurement by \textit{Planck}, $\tau=0.0544\pm 0.0073$ \citep{aghanim2020planck}, which implies a late reionization completed at $z\sim 5.5$. In the refined model, for each halo that meets the Pop~III SF criteria, we delay Pop~III SF by a collapse timescale $t_{\rm col}=\max[10^{\sigma_{\rm col}}t_{\rm dyn},\rm 3\ Myr]$. Here $t_{\rm dyn}= t_{H}/10^{3/2}$, given the Hubble time $t_{H}$, is an estimation of the free-fall timescale in the inner region ($r\lesssim 0.1 R_{\rm vir}$, given the halo virial radius $R_{\rm vir}$) where the star-forming cloud resides \citep{bromm2002}. $\sigma_{\rm col}$ is a random number generated from a uniform distribution in [-0.15,0.15], introduced to capture the diversity in halo assembly histories and avoid spurious fluctuations in the Pop~III SF rate density (SFRD). {The probability distribution of $\sigma_{\rm col}$ is chosen (by trial-and-error) to reproduce the SFRD from the cosmological simulation in \citet[see below]{campisi2011population}.} We also apply a lower bound of $10^{6}\ \rm M_{\odot}$ to the critical halo mass threshold for Pop~III SF, considering the effect of baryon-dark matter streaming motion \citep{Anna2018}. 

To keep track of Pop~III binary remnants in structure formation, we design a scheme of sampling binary stars in Pop~III star-forming haloes, based on N-body simulations of Pop~III star clusters from \citet{liu2021binary}. Following the fiducial model in \citet{liu2021binary}, we assume the same Pop~III IMF with a log-flat form ($dN_{\star}/dm_{\star}\propto m_{\star}^{-1}$) in the range between $m_{\min}=1\ \rm M_{\odot}$ and $m_{\max}=170\ \rm M_{\odot}$. In this model, the average fraction of stellar mass in binaries is $f_{\rm B}=0.69$. {The mass range adopted here, $m_{\star}\sim 1-170\ \rm M_{\odot}$, is generally consistent with that inferred from the MW stellar metallicity distribution function in the low-metallicity regime, $m_{\star}\sim 2-180\ \rm M_{\odot}$ \citep{tarumi2020}.} For each Pop~III star-forming halo, the mass $M_{\rm III}$ and number $N_{\star}$ of Pop~III stars are given by the IMF sampling scheme in HT15 for individual stars. If $N_{\star}>1$, we sample binary stars until the total mass in binaries exceeds $f_{\rm os}f_{\rm B}M_{\rm III}$, or the number of stars left is less than 2. Here, we introduce $f_{\rm os}\simeq 0.3$ to suppress overshooting in our numerical sampling, determined by trial-and-error, to make sure that the total mass of Pop~III binaries (ever formed) agrees with the total mass of Pop~III stars multiplied by $f_{\rm B}$. For each binary, we first generate the primary mass $m_{\star,1}$ from a uniform distribution in [$m_{\min}$,$m_{\rm max}$], and then generate the secondary mass from a uniform distribution in [$m_{\min}$,$m_{\star,1}$]. The mass distributions adopted here are meant to reproduce those in N-body simulations (see fig.~9 and 10 in \citealt{liu2021binary})\footnote{Here we have ignored the effect of stellar collisions, as they are rare ($\lesssim 1\%$) in most cases \citep{liu2021binary}.}, which reflect the fact that massive stars are favored for binaries. 

Following \citet[see their fig.~1]{bl2020gw190521}, we map binary stars to binary remnants with the remnant-initial mass relation $m_{\rm rem}(m_{\star})$ based on the fitting formula from \citet{tanikawa2020fitting} for their $Z=10^{-6}\ \rm Z_{\odot}$ models\footnote{In this case, NSs and BHs ($m_{\rm rem}\gtrsim 1.4\ \rm M_{\odot}$) are formed at $m_{\star}\gtrsim 10\ \rm M_{\odot}$. Stars with $m_{\star}\sim 25-85\ \rm M_{\odot}$ collapse directly into BHs. The gap for (pulsational) pair-instability supernovae is $\sim115-230\ (85-115)\ \rm M_{\odot}$.}, taking into account PPISNe: $m_{\star,i}\rightarrow m_{i}\equiv m_{\rm rem}(m_{\star,i})$ ($i=1,\ 2$)\footnote{We have ignored the effects of mass transfer and supernova explosions on the survival and properties of binary remnants. It is found in \citet{liu2021binary} that mass transfer from Roche lobe overflow is rare for Pop~III binaries. It will be shown in Sec.~\ref{s3.2.2} (Fig.~\ref{mdis}) that most mergers of Pop~III remnants involve BHs with $m_{\rm rem}\gtrsim 25\ \rm M_{\odot}$, which tend to form directly without supernova explosions.}. 
As we are concerned with the GWs from binary compact object merges, we only keep track of the remnant binaries with progenitor masses $m_{\star,i}>10\ \rm M_{\odot}$ ($i=1,\ 2$), and total masses $m_{\bullet}\equiv m_{1}+m_{2}> 3\ \rm M_{\odot}$. On average, $\sim 10^{4}$ such binaries are formed in each tree. 
It will be shown below that the distribution of total mass $m_{\bullet}$ in remnant binaries obtained by our sampling scheme is generally consistent with that derived from the binary statistics of N-body simulations in \citet{liu2021binary}. 

\subsubsection{Pop~I/II star formation}
\label{s2.1.2}

\begin{figure}
    \centering
    \includegraphics[width=1\columnwidth]{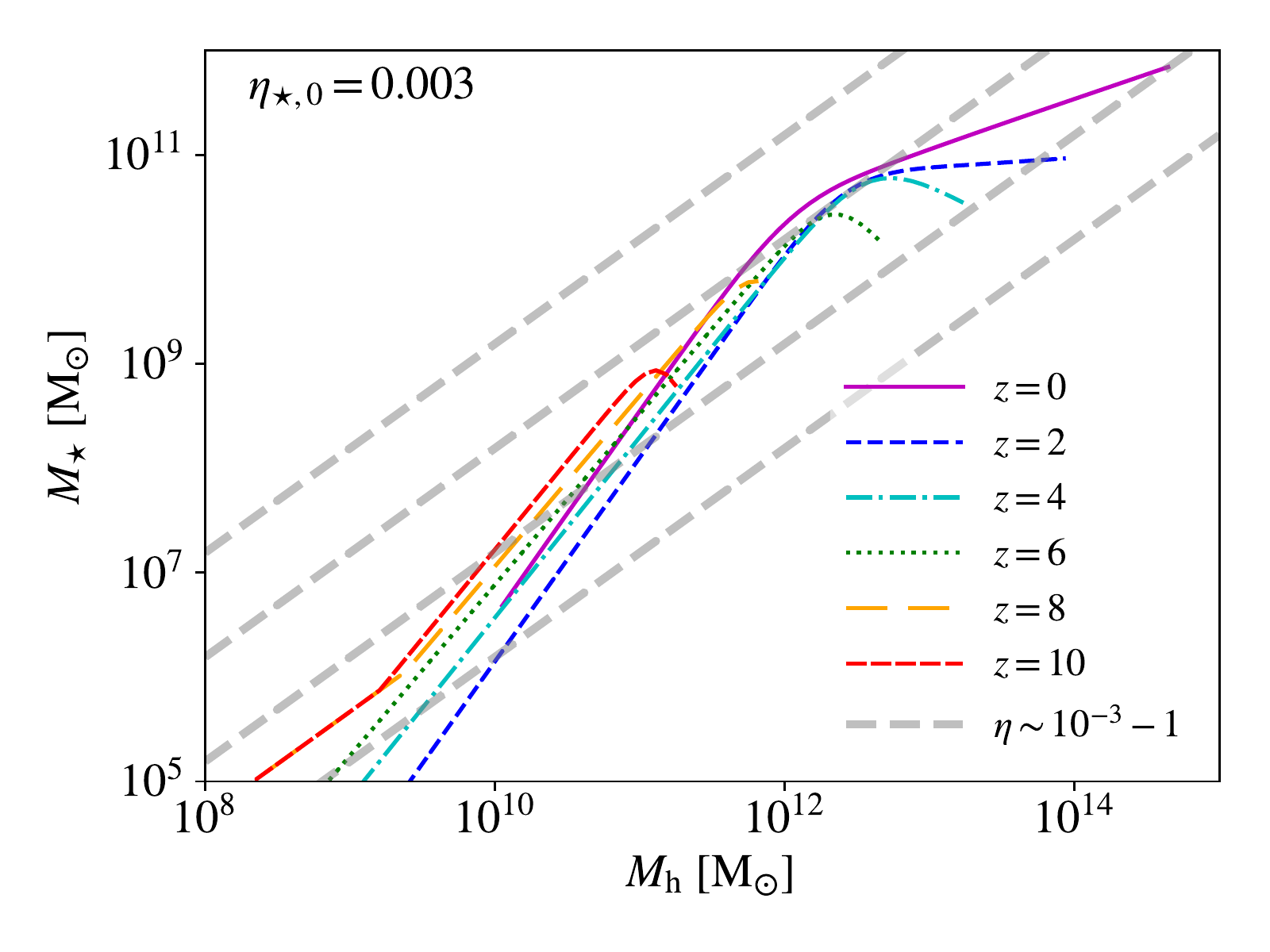}
    \vspace{-25pt}
    \caption{SHMR from \citet{behroozi2019universemachine} at $z=0$ (solid), 2 (dashed), 4 (dashed-dotted), 6 (dotted), 8 (long-dashed) and 10 (nest-dashed), given the (pre-reionization) lower limit of SFE $\eta_{\star,0}=0.003$. The thick dashed lines indicate constant SFEs in the range of $10^{-3}-1$.}
    \label{msmh}
\end{figure}

For simplicity, Pop~I/II SF is modelled with an input SFRD in HT15 to calculate the relevant contribution to the Thomson optical depth, while not explicitly taking into account the chemical and radiative feedback from Pop I/II stars. In our case, we need to derive the properties of individual galaxies that potentially host NSCs in order to keep track of the dynamics of Pop~III (binary) remnants. Therefore, we now include a self-consistent model for Pop~I/II SF and feedback based on the stellar-halo mass relation (SHMR) in \citet{behroozi2019universemachine}. We adopt their fitting formula for the true mass values including both star-forming and quiescent galaxies and excluding intrahalo light (see column 8 of their table J1), as shown in Fig.~\ref{msmh}. We further impose a lower limit for the SF efficiency (SFE)\footnote{Ratio of the stellar to total baryon mass.}, $\eta_{\star,0}$, in (the extrapolation of) this relation before reionization ($z>6$), which is particularly relevant for high-$z$ haloes with relatively low masses ($M_{\rm h}\lesssim 10^{9}\ \rm M_{\odot}$). We assume that all haloes enriched by metals with $M_{\rm h}>10^{6}\ \rm M_{\odot}$ can host Pop I/II stars. The stellar masses of such haloes are assigned and updated according to the SHMR: $M_{\star}\equiv M_{\star}(M_{\rm h},z)$. For each new (existing) enriched halo, (the increase of) $M_{\star}$ (with respect to the stellar mass carried by child haloes at the previous timestep) is attributed to newly-formed Pop~I/II stars at the current timestep. To take into account the effect of reionization, we prevent any halo with $M_{\rm h}<6.7\times 10^{8}\ \mathrm{ M_{\odot}}\ [(1+z)/5]^{-3/2}$ from forming new stars at $z<6$ based on reionzation models in \citet{pawlik2015spatially,pawlik2017aurora,benitez2020detailed,hutter2020astraeus}. 

For chemical feedback from Pop~I/II stars, we simply increase the volume filling fraction of metals from Pop~III enrichment by a factor of 5 to mimic the additional enrichment from galactic outflows of Pop~I/II galaxies formed in regions initially enriched by Pop~III stars. While for LW feedback, we add the contribution from Pop~I/II stars to the LW background (in units of
$10^{-21}\ \rm erg\ s^{-1}\ cm^{-2}\ Hz^{-1}\ sr^{-1}$) following \citet{johnson2013first}:
\begin{align}
    J_{\rm 21,II,bg}=0.3\left(\frac{1+z}{16}\right)^{3}\left[\frac{{\rm SFRD_{\rm I/II}}(z)}{10^{-3}\ \rm M_{\odot}\ yr^{-1}\ Mpc^{-3}}\right]\ ,
\end{align}
where ${\rm SFRD_{\rm I/II}}(z)$ is the \textit{co-moving} SF rate density of Pop~I/II stars self-consistently predicted by the above model, based on the SHMR.

\begin{figure}
    \centering
    \includegraphics[width=1\columnwidth]{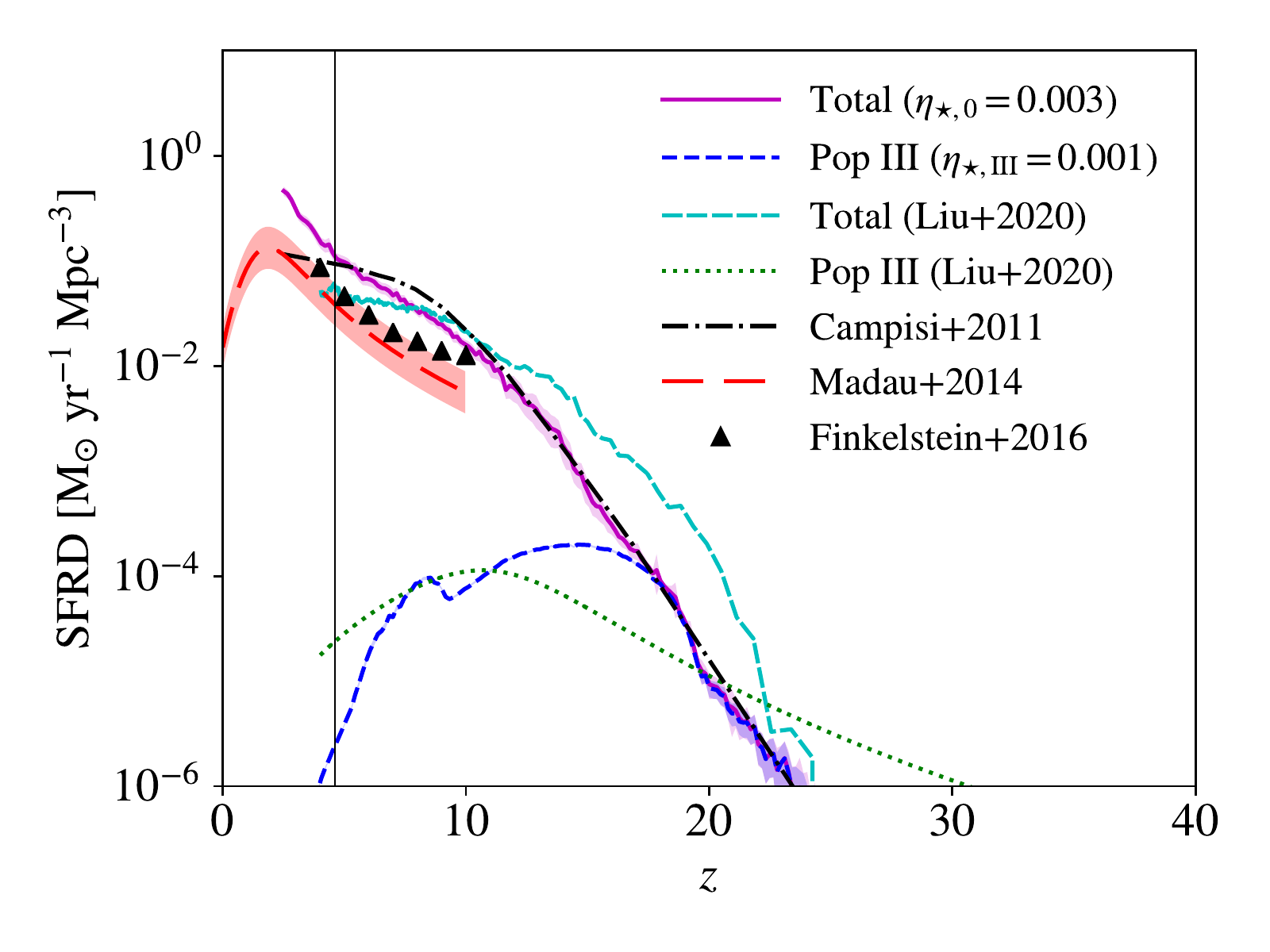}
    \vspace{-25pt}
    \caption{Co-moving SFRDs for all stars (solid) and Pop~III stars (dashed) with $\eta_{\star,0}=0.003$ and $\eta_{\star,\rm III}=0.001$. $3\sigma$ uncertainties are illustrated with the shades. The reference total SFRD from \citet{campisi2011population} is shown with the dashed-dotted curve. The observational results in \citet{madau2014araa}, inferred from UV and IR galaxy surveys such as \citet[data points]{finkelstein2016observational}, are plotted as the long-dashed curve (with scatters of 0.2 dex embodied by the shaded region). For comparison, we also shown the total and Pop~III SFRDs from \citet{liu2020gw,liu2020did} with the nest-dashed and dotted curves, respectively. The thin vertical line denotes the turn-over redshift $z\simeq 4.6$ of the target halo below which our results may not be cosmologically representative.}
    \label{sfrd}
\end{figure}

\begin{figure}
    \centering
    \includegraphics[width=1\columnwidth]{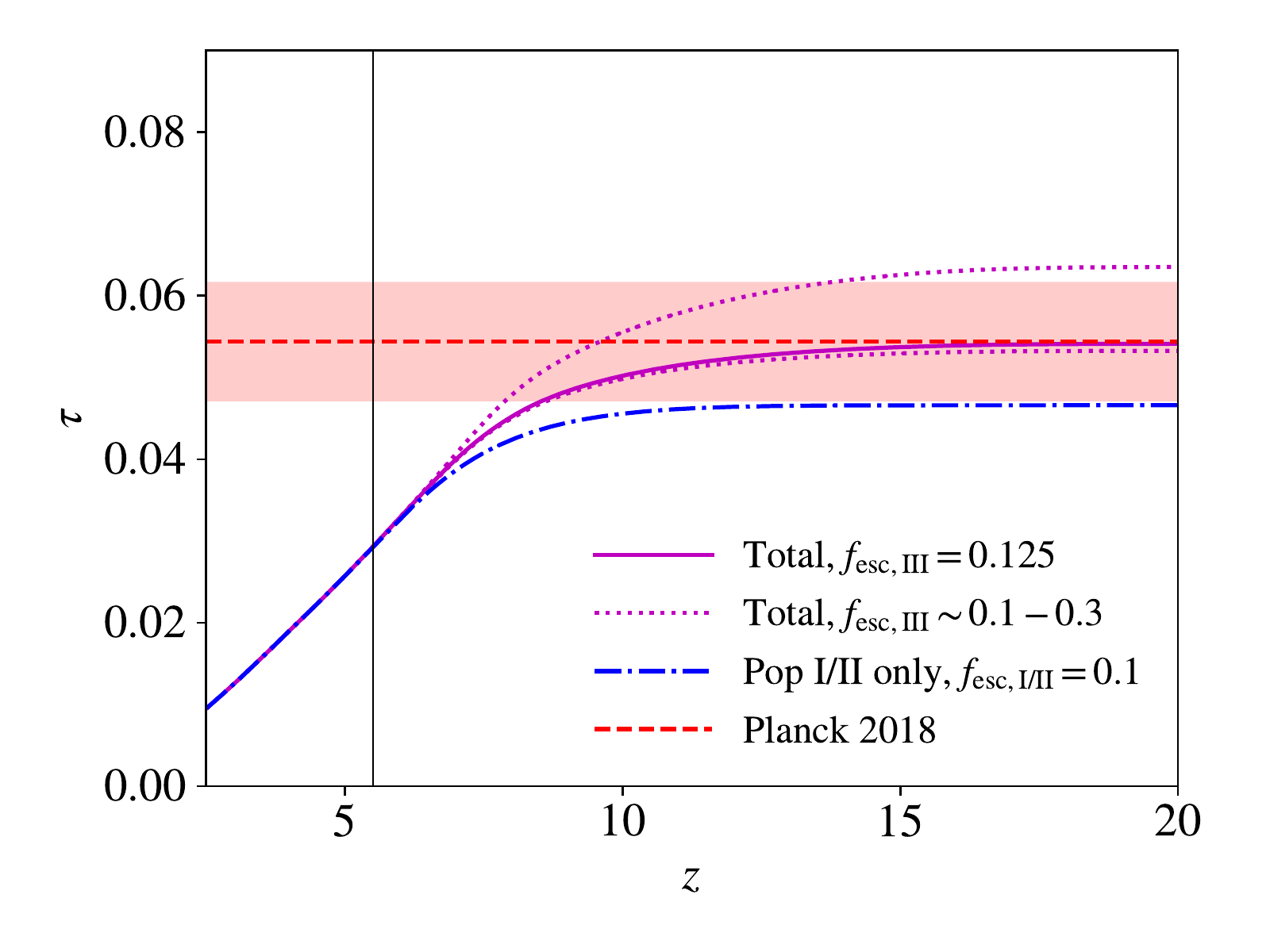}
    \vspace{-25pt}
    \caption{Thomson optical depth as a function of redshift. The results with all contributions from Pop~III and I/II stars are shown with the solid and dotted curves, assuming a Pop~I/II escape fraction of $f_{\rm esc,I/II}=0.1$, and considering a range of Pop~III escape fractions of $f_{\rm esc,III}\sim 0.1-0.3$, where 0.125 is the best fit value (solid). The results with only Pop~I/II stars are shown with the dashed-dotted curve (again for $f_{\rm esc,I/II}=0.1$). The recent measurement by \textit{Planck} \citep{aghanim2020planck} is shown with the horizontal dashed line, whose $1\sigma$ scatters are denoted by the shaded region. For reference, $z=5.5$ is shown with the thin vertical line.}
    \label{tau}
\end{figure}

Finally, we choose the (pre-reionization) lower limit for the Pop~I/II SFE, $\eta_{\star,0}$, and the Pop~III SFE, $\eta_{\rm III}$, to reproduce the total (Pop~III+Pop~I/II) SFRD inferred from observations and simulations. Following HT15, we take the results from the cosmological simulation in \citet{campisi2011population} as the reference, which are consistent with the observational results in \citet{madau2014araa} at $z\lesssim 4$ where observational constraints are most reliable. As shown in Fig.~\ref{sfrd}, the reference SFRD can be well reproduced with $\eta_{\star,0}=0.003$ and $\eta_{\star,\rm III}=0.001$, particularly for $z\gtrsim 5$ when the target halo has not reached turn-over. We also compare our SFRDs with those derived from the recent cosmological simulation \texttt{FDbox\_Lseed} in \citet{liu2020gw,liu2020did}. Our results are generally consistent with simulation data within an order of magnitude, especially around the peak of Pop~III SF at $z\sim 6-12$. It will be shown below that this is also the peak of NSC capture of Pop~III remnants (Sec.~\ref{s3.1}). Note that the early cosmic SF history is still highly uncertain, particularly for Pop~III stars with very limited constraints from observations and significant difference in the predictions from theoretical studies with different numerical models and assumptions for SF and stellar feedback (see e.g. fig. 5 and 13 of \citealt{liu2020did}). Hopefully more robust observational constraints can be obtained in the \textit{JWST} era. We defer more detailed calibrations with SFRD to future work. Instead, when comparing our predicted MRD of Pop~III remnants with literature results, we constrain the Pop~III SF history by scaling their SFRDs (and MRDs) such that the scaled integrated Pop~III stellar mass densities ($\mathrm{ISMD}_{\rm III}\equiv\int_{0}^{\infty}\mathrm{SFRD}_{\rm III}|dt/dz|dz$) are identical to our value $\simeq 7\times 10^{4}\ \mathrm{M_{\odot}\ Mpc^{-3}}$.

Actually, the Thomson optical depth measured by \textit{Planck}, $\tau=0.0544\pm 0.0073$ \citep{aghanim2020planck}, can also be well reproduced by our SF model under reasonable assumptions for the escape fractions of ionizing photons. Fig.~\ref{tau} shows the evolution of $\tau$ with redshift predicted by our merger trees, which agrees well with the \textit{Planck} result assuming escape fractions $f_{\rm esc,II}=0.1$ for Pop~I/II and $f_{\rm esc,III}\sim 0.1-0.3$ for Pop~III. With these escape fractions, the predicted volume filling fraction of ionized gas reaches $\simeq 1\ (0.5)$ at $z\simeq 5.5\ (8)$, consistent with the picture of late reionization. The escape fractions adopted here are within the typical range $f_{\rm esc}\sim 0.1-0.7$ seen in simulations and semi-analytical models (e.g. \citealt{so2014fully,paardekooper2015first,inayoshi2016gravitational,visbal2020self,katz2021introducing}). 

\subsection{In-spiral of Pop~III remnants by dynamical friction}
\label{s2.2}

Once formed in small structures, Pop~III (binary) remnants\footnote{We ignore the growth of Pop~III remnants through accretion, as previous studies based on 3D cosmological simulations have found that stellar-mass Pop~III remnants can hardly grow via accretion at high-$z$ (e.g. \citealt{johnson2007aftermath,alvarez2009accretion,hirano2014one,smith2018growth}).} will fall into larger structures during halo mergers and meanwhile spiral towards galaxy centres by DF from field stars\footnote{For simplicity, we ignore the DF from gas, whose effect is likely minor in the long-term \citep{chen2021dynamical}. We also ignore the effect of dark matter, as we are only concerned with the central region ($r\lesssim 300\ \rm pc$) dominated by baryons in most cases, where the DF timescale is shorter than the Hubble time.}, where they can further fall into NSCs. To model this process, we need to consider how remnants are (initially) distributed in haloes during structure formation, as well as the host galaxy properties. 

\begin{figure}
    \centering
    \includegraphics[width=1\columnwidth]{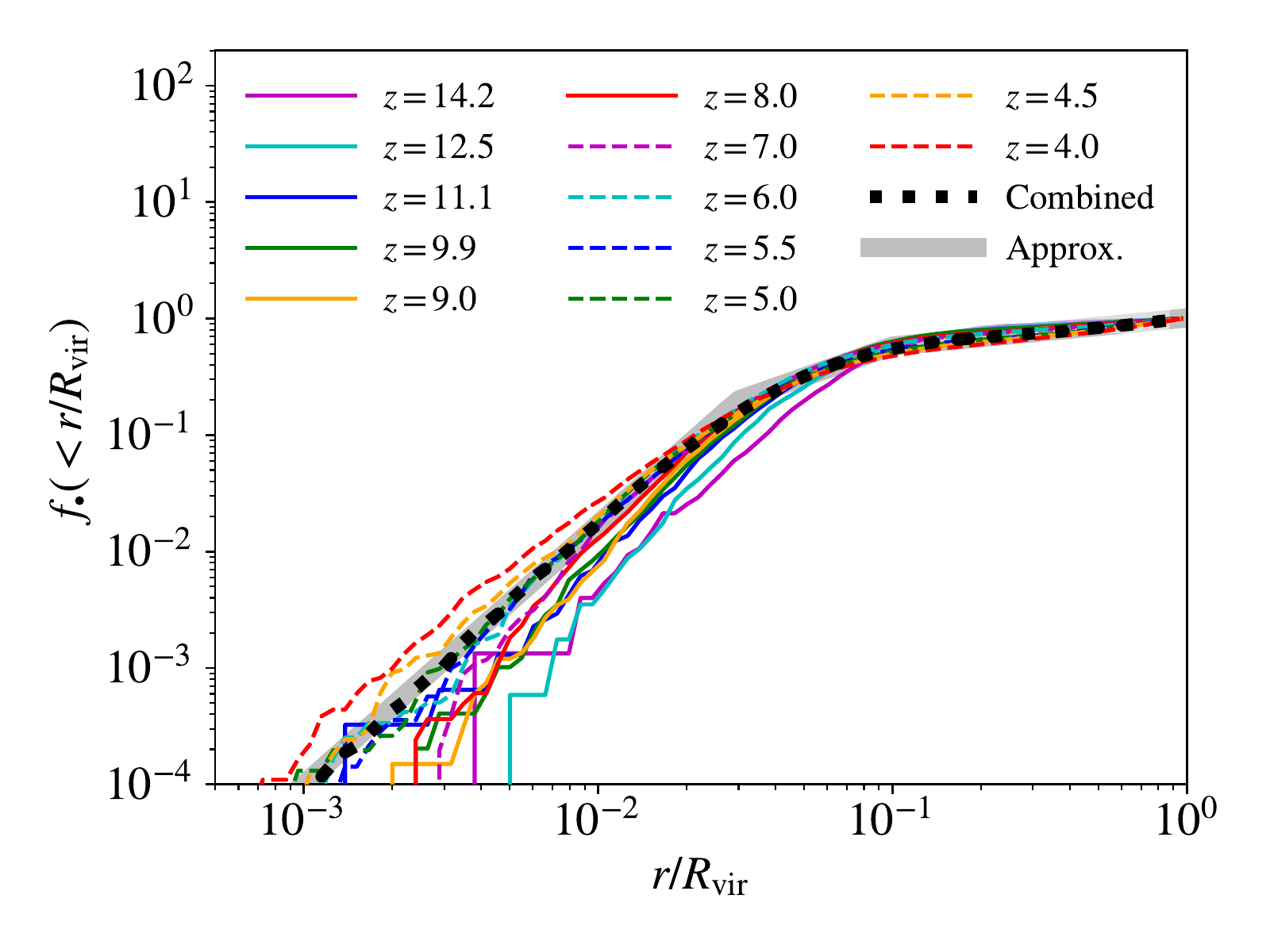}
    \vspace{-25pt}
    \caption{Enclosed fractions of Pop~III remnants in terms of $r/R_{\rm vir}$ given the physical distance $r$ to the galaxy/halo centre and the halo virial radius $R_{\rm vir}$, for 12 snapshots at $z\sim 4-14.2$ from the cosmological simulation in \citet{liu2020gw,liu2020did}. The combined distribution for all snapshots is shown with the dotted curve, which is approximated with a triple power-law fit (thick solid) in our numerical model. The inner power-law slope is $\sim 2$.}
    \label{fbhbrv}
\end{figure}

For the former, we adopt the spatial distribution of remnants from the cosmological simulation in \citet{liu2020gw,liu2020did}, as shown in Fig.~\ref{fbhbrv}. The distribution has little evolution at $z\sim 4-14$ in units of $r/R_{\rm vir}$, where $r$ is the physical distance to the galaxy/halo centre and $R_{\rm vir}$ is the halo virial radius. All newly-formed Pop~III remnants are distributed in their initial hosts following this distribution. When two haloes merge, the remnants at $r>3\ \rm pc$ in the smaller progenitor are distributed into the merged halo with the same distribution, while the locations/orbits of the remnants in the larger progenitor remain unaffected by the merger.

For the latter, we adopt the galaxy model in \citet[AC14, henceforth]{arca2014globular}, which successfully reproduces several observed scaling relations between NSCs and host galaxies for the globular cluster (GC) in-spiral scenario. In this model, a galaxy is characterized by a Dehnen sphere with density profile \citep{dehnen1993}
\begin{align}
    \rho_{\gamma}(r|M,R,\gamma)=\frac{\rho_{\gamma 0}}{(r/R)^{\gamma}(1+r/R)^{4-\gamma}}\ ,\label{e2}
\end{align}
where $\rho_{\gamma 0}=(3-\gamma)M/(4\pi R^{3})$, given $M$, $R$ and $\gamma$ as the total mass, scale length and inner slope. For each galaxy\footnote{Each halo labelled as enriched by the stochastic enrichment model in HT15 hosts a galaxy. To take into account self enrichment, we further assign a galaxy to each halo that has hosted Pop~III SF even if it is not enriched according to the stochastic model. The (Pop~I/II) galaxy formation in such self-enriched haloes is delayed by $100\ \rm Myr$ with respect to the initial Pop~III SF event, which is the typical re-collapse timescale after supernovae.}, the galaxy (stellar) mass $M_{\star}$ is given by our Pop~I/II SF model. We then estimate the galaxy size (scale length) $R_{\rm g}$ with the size-mass relation (AC14)
\begin{align}
    R_{\rm g}/\mathrm{kpc}=2.37\ [2^{1/(3-\gamma_{\rm g})}-1]M_{\rm \star,11}^{0.14}\ ,\label{e3}
\end{align}
where $M_{\star,11}\equiv M_{\star}/(10^{11}\ \rm M_{\odot})$, and the galaxy inner slope $\gamma_{\rm g}$ is generated from a uniform distribution in the ranges of $0-0.5$ for $M_{\star,11}<0.1$ and $0.5-1$ for $M_{\star,11}\ge 0.1$ (for newly-formed galaxies). During halo mergers, $\gamma_{\rm g}$ is inherited from the larger progenitor (i.e. along the main branch). Since we are concerned with high-$z$ low-mass galaxies, we always have $\gamma_{\rm g}\sim 0-0.5$. Considering a broader range $\gamma_{\rm g}\sim 0-2$ hardly changes our results. 

Now that the initial conditions and galaxy backgrounds are set up, we can estimate the DF timescale, $\tau_{\rm DF}$, for remnants to experience their in-spiral. For an object of mass $m$ with initial apocentric distance $r$ and velocity $v$ with respect to the centre in a system of total mass $M$ and size $R$, the DF timescale given by Chandrasekhar's formula is \citep{bt2008}
\begin{align}
    \frac{\tau_{\rm DF}}{\mathrm{Myr}}=\frac{34.2}{\ln\Lambda}\left(\frac{r}{\rm3\ pc}\right)^{2}\left(\frac{v}{\rm 10\ km\ s^{-1}}\right)\left(\frac{m}{100\ \rm M_{\odot}}\right)^{-1}\ , \label{e4}
\end{align}
where $\ln\Lambda\sim\ln[Mr/(0.8mR)]$ is the Coulomb logarithm, and the initial velocity is estimated with the velocity dispersion $v\sim\sigma\sim\sqrt{GM/R}$. 

The work of Chandrasekhar has been generalized to describe DF in both cusped and cored density profiles by \citet{arca2015henize,arca2016formation} which give an updated DF timescale formula for the Dehnen profile (Equ.~\ref{e2}):
\begin{align}
    \frac{\tau_{\rm DF}}{\mathrm{Myr}}=0.3g\sqrt{\frac{r^{3}}{\rm 1\ kpc^{3}}M_{11}^{-1}}\left(\frac{m}{M}\right)^{\alpha}\left(\frac{r}{R}\right)^{\beta}\ ,\label{e5}
\end{align}
where $M_{11}=M/(10^{11}\ \rm M_{\odot})$, $\alpha=-0.67$, $\beta=1.76$, and 
\begin{align}
    g&\equiv g(e_{\rm if},\gamma)\notag\\
    &=(2-\gamma)\left\{a_{1}\left[\frac{1}{(2-\gamma)^{a_{2}}}+a_{3}\right](1-e_{\rm if})+e_{\rm if}\right\}\ .
\end{align}
Here, $e_{\rm if}$ is the eccentricity of the in-fall orbit, $a_{1}=2.63\pm0.17$, $a_{2}=2.26\pm 0.08$, and $a_{3}=0.9\pm 0.1$. Note that this formula is only valid for $0<\gamma<2$. In our implementation, we take the \textit{minimum} of the above two DF timescales, which are coupled to our galaxy model via $m=m_{\bullet}$, $M=M_{\star}$, $R=R_{\rm g}$ and $\gamma=\gamma_{\rm g}$. Actually, in most cases, the recent formula~(\ref{e5}) gives a shorter DF timescale. 

\begin{figure}
    \centering
    \includegraphics[width=1\columnwidth]{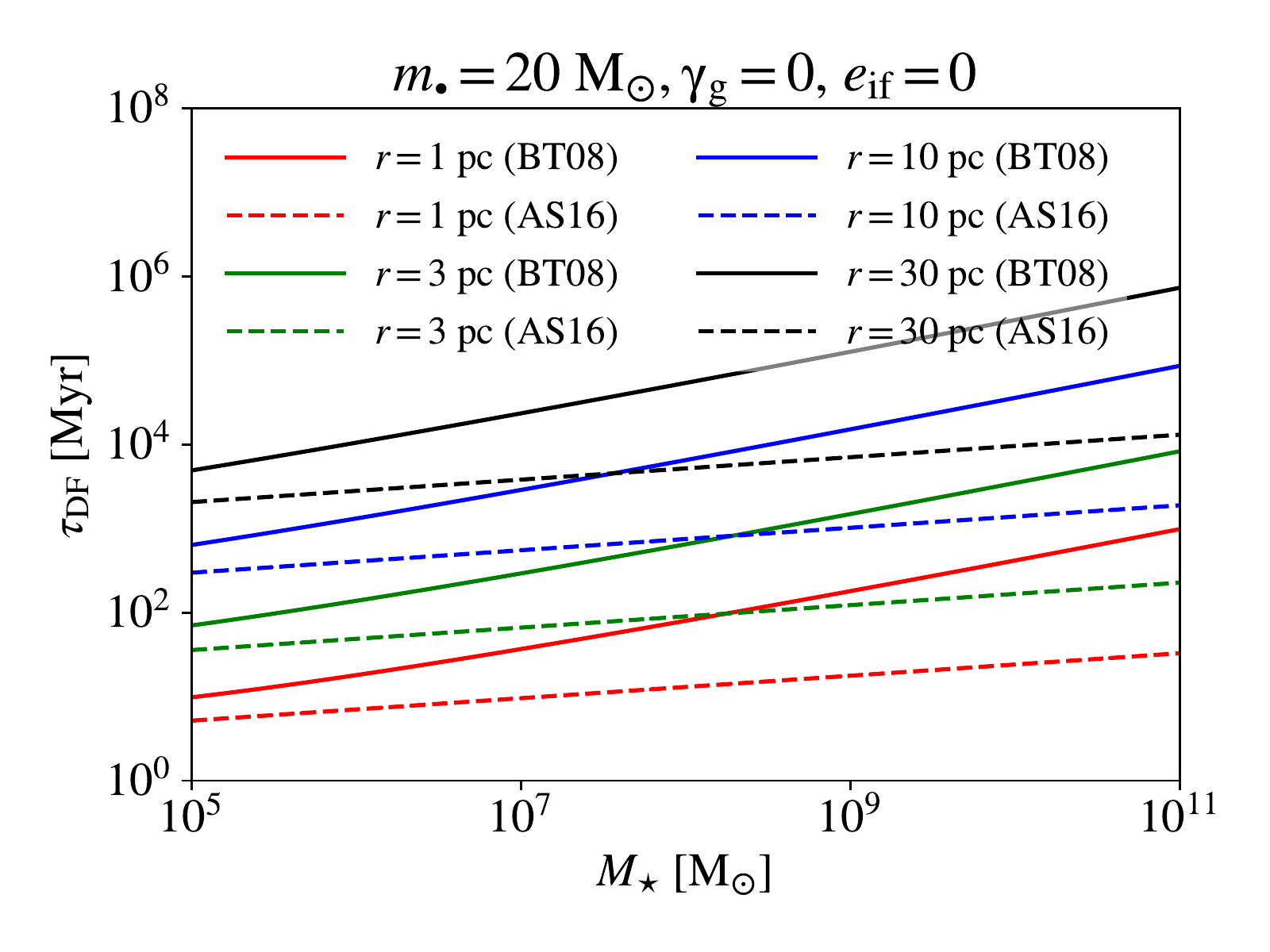}
    \vspace{-25pt}
    \caption{DF timescale as a function of galaxy mass, for $m_{\bullet}=20\ \rm M_{\odot}$, $\gamma_{\rm g}=0$ and $e_{\rm if}=0$, at $r\sim 1,\ 3,\ 10\ \text{and}\ 30\ \rm pc$ (from bottom to top), given by the Chandrasekhar's formula \citep[BT08 solid, see Equ.~\ref{e4}]{bt2008}, and more recent work in \citet[AS16, dashed, see Equ.~\ref{e5}]{arca2016formation}. }
    \label{tdf}
\end{figure}

Fig.~\ref{tdf} demonstrates the dependence of $\tau_{\rm DF}$ on $M_{\star}$ for $m_{\bullet}=20\ \rm M_{\odot}$, $\gamma_{\rm g}=0$ and $e_{\rm if}=0$. It turns out that smaller galaxies are more efficient at DF. For most binaries with $m_{\bullet}\gtrsim 20\ \rm M_{\odot}$ in our case, we have $\tau_{\rm DF}\lesssim 100\ \rm Myr$ for $r\lesssim 3\ \rm pc$, shorter than the typical timescale for galaxy evolution, and $\tau_{\rm DF}\gtrsim 10\ \rm Gyr$ for $r\gtrsim 300\ \rm pc$, comparable/longer than the Hubble time. Therefore, we only compute the insprial of remnants in the distance range $r\sim 3-300\ \rm pc$. The evolution of $r$ is described by the differential equation
\begin{align}
    \frac{dr}{dt}\simeq-\frac{r}{\tau_{\rm DF}(r|m,M,R,\gamma,e_{\rm if})}\ ,\label{e7}
\end{align}
with $m=m_{\bullet}$, $M=M_{\star}$, $R=R_{\rm g}$ and $\gamma=\gamma_{\rm g}$, where $M_{\star}$ and $R_{\rm g}$ are updated on-the-fly. As a conservative choice, we set $e_{\rm if}=0$. We have verified that considering higher $e_{\rm if}$ up to 0.9 only slightly increases the fraction of remnants with $r<3\ \rm pc$. We record all remnants that fall into galaxy centres ($r<3\ \rm pc$), together with their host properties. For remnants within the target halo at $z\simeq 2.5$, we further evolve them to $z=0$ by post-processing with galaxy properties fixed at $z\simeq 2.5$, and record the ones that will fall into the centre at $z>0$. This offers a rough estimation for the low-redshift regime, which will be explored in future work. %

\begin{figure}
    \centering
    \includegraphics[width=1\columnwidth]{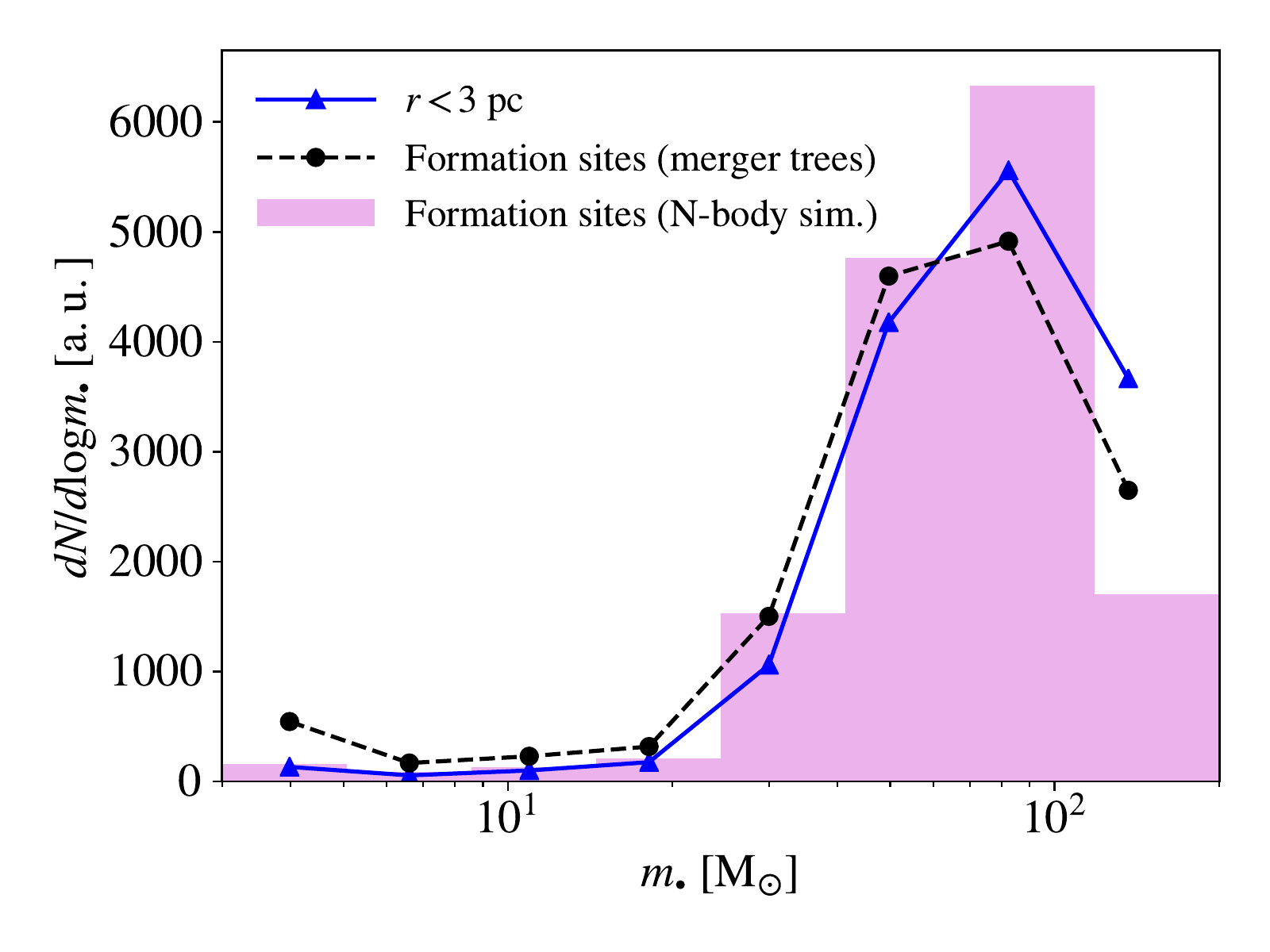}
    \vspace{-25pt}
    \caption{Mass distributions of Pop~III binary remnants at galaxy centres ($r<3\ \rm pc$, solid) as well as formation sites, from our binary star sampling scheme (dashed) and the N-body simulations in \citet[histograms]{liu2021binary}.}
    \label{bhbmsamp}
\end{figure}

\begin{figure}
    \centering
    \includegraphics[width=1\columnwidth]{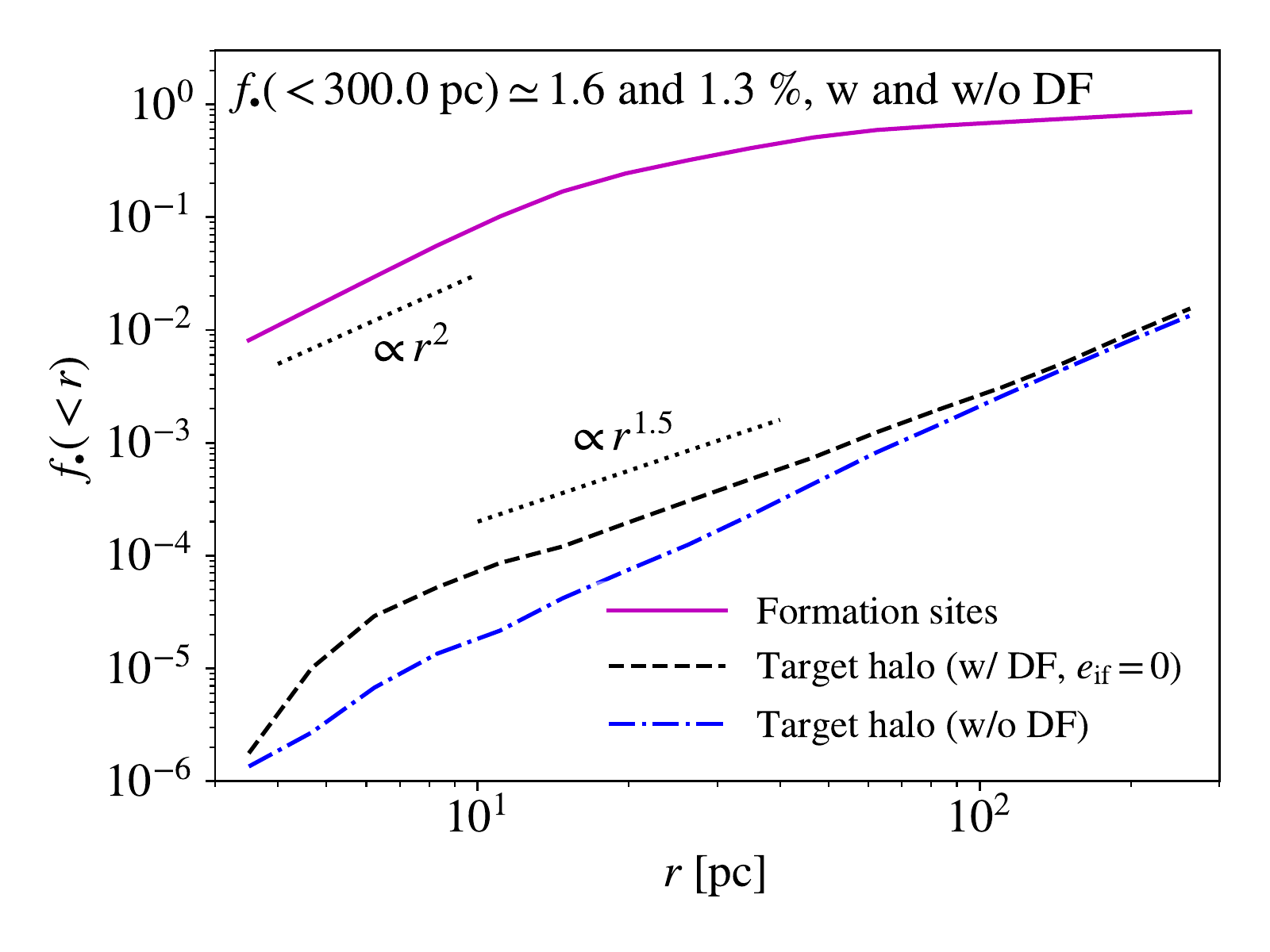}
    \vspace{-25pt}
    \caption{Enclosed fractions of Pop~III binary remnants in terms of the physical distance $r$, at formation sites (solid) and the target halo with (dashed) and without (dashed-dotted) DF. All distributions are normalized to the total number of binary remnants formed in the merger trees. Without DF, the inner distribution follows $f_{\bullet}(<r)\propto r^{2}$ approximately (dotted), at both the target halo and formations sites, consistent with the input simulation data (Fig.~\ref{fbhbrv}).}
    \label{bhbloc}
\end{figure}

On average, 7.5\% of Pop~III remnants fall into the centres ($r<3\ \rm pc$) of galaxies with $M_{\rm NSC}>10^{5}\ \rm M_{\odot}$ in our merger trees. These remnants can further fall into NSCs, depending on the occupation frequency of NSCs (see below). As DF is more efficient for more massive in-falling objects, the mass distribution of remnants with $r<300\ \rm pc$ is more top-heavy compared with the initial mass distribution at formation sites, as shown in Fig.~\ref{bhbmsamp} in arbitrary units (a.u.). It is also shown that the mass distribution obtained by our sampling scheme is generally consistent with that derived from the N-body simulations in \citet{liu2021binary}, although the most massive ($m_{\bullet}\gtrsim 100\ \rm M_{\odot}$) and low-mass ($m_{\bullet}\lesssim 30\ \rm M_{\odot}$) objects are slightly overproduced. 

To further illustrate the dynamics of Pop~III remnants in their host galaxies during structure formation, Fig.~\ref{bhbloc} shows the spatial distributions of Pop~III binary remnants at formation sites and the target halo. In general, due to cosmic expansion and the nature of DF (see Fig.~\ref{tdf}), it is more difficult for Pop~III remnants to enter inner regions of galaxies in more massive haloes at later times. In the formation sites with $M_{\rm h}\sim 10^{6-7}\ \rm M_{\odot}$ at $z\sim 5-20$, $\sim 90$\% of remnants are initially within $300\ \rm pc$, while only $\sim 1-2$\% of remnants can fall into the inner region ($r\lesssim 300\ \rm pc$) of the target halo ($M_{\rm h}\sim 10^{12}\ \rm M_{\odot}$ at $z\simeq 2.5$). DF slightly increases the fraction in the target halo but the effect is rather minor. This implies that NSCs formed in low-mass galaxies at high-$z$ will be the dominant hosts of Pop~III binary remnants, which is confirmed by our results (see Sec.~\ref{s3.1}).

\subsection{Evolution of compact object binaries in NSCs}
\label{s2.3}

\begin{figure}
    \centering
    \includegraphics[width=1\columnwidth]{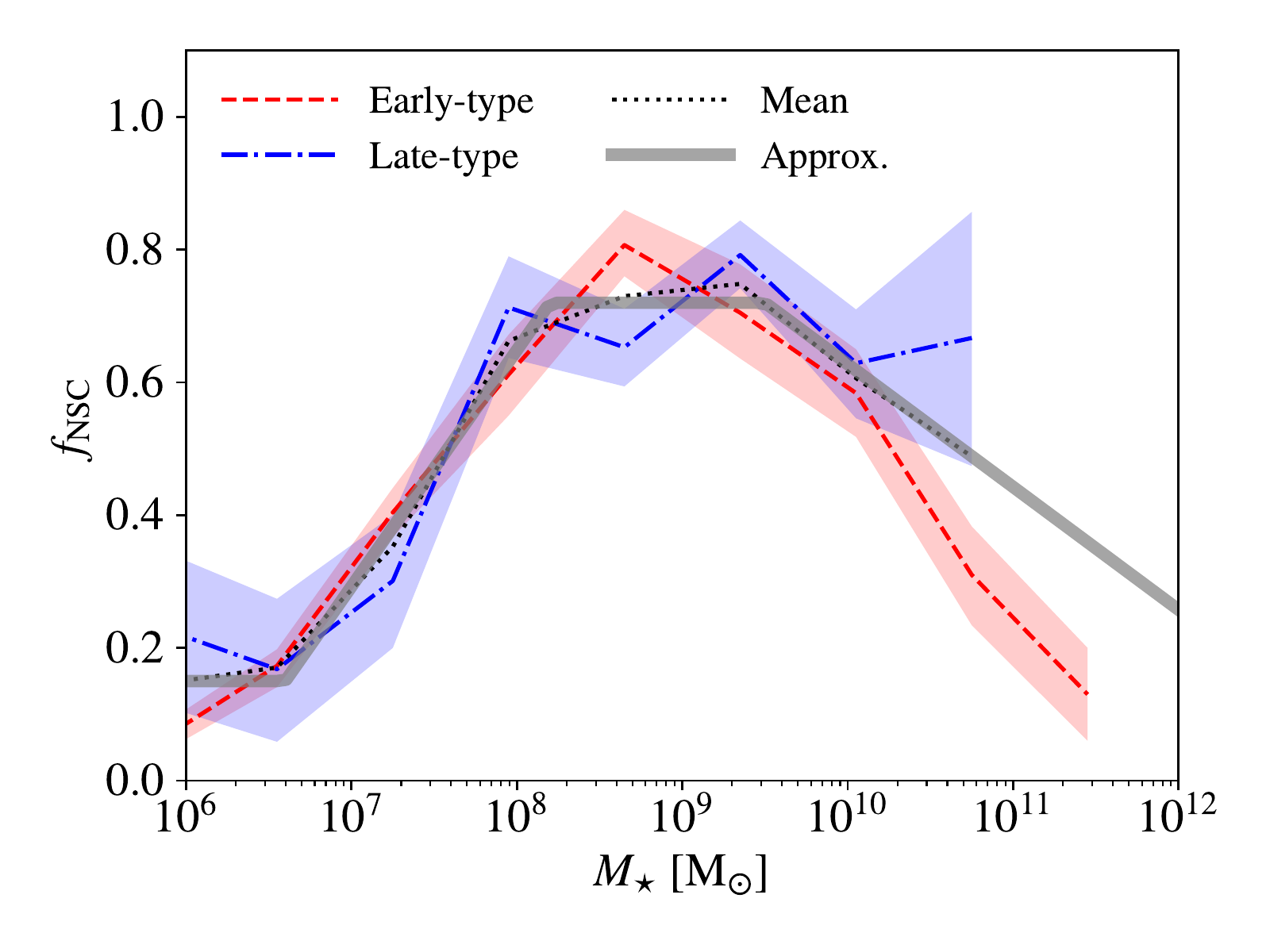}
    \vspace{-25pt}
    \caption{Occupation frequency of NSCs as a function of galaxy mass for early-type (dashed) and late-type (dashed-dotted) galaxies in observations (\citealt{neumayer2020nuclear}, see their fig~3). The dotted curve shows the mean, combining the two types of galaxies assuming a 1:1 mixture ratio, as observed in the general field \citep{calvi2012distribution}. The thick solid curve is our approximation to the observed mean, in which $f_{\rm NSC}=0.15$ at the low-mass end.}
    \label{foccnsc}
\end{figure}

For the $7.5\%$ of Pop~III binary remnants at galaxy centres ($r<3\ \rm pc$), a fraction of them will further fall into NSCs\footnote{We ignore the binary evolution outside NSCs since most of our binary remnants are initially wide (with separations larger than a few au) with very weak GW emission, and the density of stars in the galaxy field is too low for efficient DH.} in which their evolution is driven by DF and three-body encounters from low-mass stars, as well as GW emission. We follow this evolution with a NSC model based on (local) observations \citep{neumayer2020nuclear} by post-processing. First, we need to estimate the occupation frequency of NSCs, $f_{\rm NSC}$, which is still uncertain for high-$z$ galaxies. We only consider NSCs with masses $M_{\rm NSC}>10^{5}\ \rm M_{\odot}$. The lower bound here is consistent with the typical mass of NSCs in the first galaxies predicted by theoretical models \citep{devecchi2009formation,devecchi2010high,devecchi2012high}, and the smallest NSCs in observations. Note that the mass of Pop~I/II star clusters in the first galaxies is still uncertain, for which lower values ($\sim 3\times 10^{3}\ \rm M_{\odot}$) are predicted by high-resolution hydrodynamic simulations (e.g. \citealt{safranek2016star}). In the optimistic case, we assume full occupation for galaxies with expected NSC masses $M_{\rm NSC}>10^{5}\ \rm M_{\odot}$ (see below). In the conservative case, we adopt the empirical occupation frequency as a function of galaxy mass, shown in Fig.~\ref{foccnsc}. In our implementation of $f_{\rm NSC}$, we have $f_{\rm NSC}=0.15$ for $M_{\star}\sim 10^{5}-3\times 10^{6}\ \rm M_{\odot}$, which is also consistent with theoretical predictions for typical host haloes for these low-mass galaxies with $M_{\rm h}\sim 10^{8}\ \rm M_{\odot}$ at $z\sim 10-20$ \citep[see their fig.~5]{devecchi2012high} and observations of dwarf galaxies.

\begin{figure}
    \centering
    \includegraphics[width=1\columnwidth]{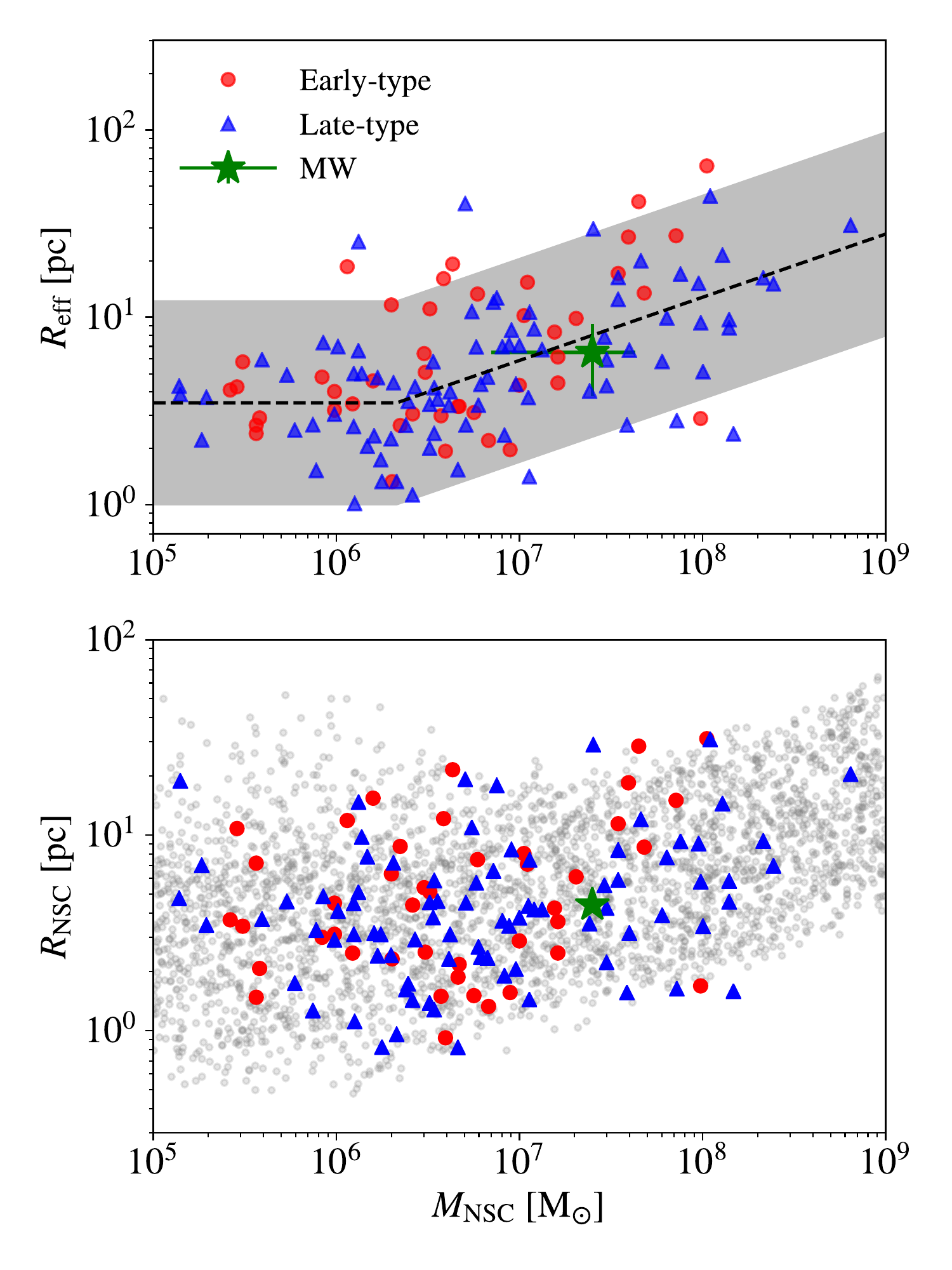}
    \vspace{-25pt}
    \caption{Size-mass relations for NSCs in terms of the effective radius (top) and scale length (bottom). Observational data compiled by \citet[see their fig. 7]{neumayer2020nuclear} are shown with circles and triangles for early-type and late-type galaxies, respectively. The MW is labelled with the green star for $M_{\rm NSC}=2.5\pm 1.8\times 10^{7}\ \rm M_{\odot}$, $R_{\rm eff}=6.5\pm 2.7\ \rm pc$ \citep{neumayer2020nuclear} and $R_{\rm NSC}\sim 4.4\ \rm pc$. In the top panel, scatters of 0.55 dex are denoted by the shaded region. In the bottom panel, typical scatters of $R_{\rm NSC}$ are shown with gray dots for a mock sample of NSCs with a log-flat mass function in the range of $M_{\rm NSC}\sim 10^{5}-10^{9}\ \rm M_{\odot}$.}
    \label{rnsc}
\end{figure}

Next, we build a stochastic model for NSC properties from empirical NSC-host scaling relations and the corresponding scatters. We again assume Dehnen profile (Equ.~\ref{e2}) for NSCs. The NSC mass $M_{\rm NSC}$ is estimated with the NSC-galaxy mass relation \citep{neumayer2020nuclear}
\begin{align}
    \log(M_{\rm NSC}/\mathrm{M_{\odot}})=0.48\log[M_{\star}/(10^{9}\ \mathrm{M_{\odot}})]+6.51\ ,
\end{align}
with scatters of $\sim 0.6$ dex. The (expected) NSC mass is assigned and updated on-the-fly for each galaxy with $M_{\star}>10^{5}\ \rm M_{\odot}$. The scatter is generated from an uniform distribution in logarithmic space, which is also inherited along the main branch. We also require $M_{\rm NSC}\le M_{\star}$. Once $M_{\rm NSC}$ is known, we derive the NSC size following the NSC size-mass relation (see fig.~7 of \citealt{neumayer2020nuclear}). We fit the observational data for effective radius $R_{\rm eff}$ and $M_{\rm NSC}$ with
\begin{align}
    \log\left( \frac{R_{\rm eff}}{\rm pc}\right)\simeq\begin{cases}
         0.54\ ,\ &M_{\rm NSC}<M_{\rm crit}\ ,\\
         0.34\log\left(\frac{M_{\rm NSC}}{\mathrm{M_{\odot}}}\right)-1.59\ ,\ &M_{\rm NSC}\ge M_{\rm crit}\ ,
    \end{cases}\notag
\end{align}
where $M_{\rm crit}\simeq 2\times 10^{6}\ \mathrm{M_{\odot}}$. The standard deviation around this fit is $\sim 0.32$ dex, which implies scatters of $\sim 0.55$ dex, assuming a uniform distribution (of scatters) in logarithmic space. For a Dehnen sphere, the relation between scale length and effective radius is (AC14)
\begin{align}
    R_{\rm NSC}=(4/3)R_{\rm eff}\left[2^{1/(3-\gamma_{\rm NSC})}-1\right]\ ,\label{e9}
\end{align}
where the inner slope $\gamma_{\rm NSC}\sim 0.65-2.55$ tends to decrease with higher $M_{\star}$ and $M_{\rm NSC}$ \citep{neumayer2020nuclear,pechetti2020luminosity}. To capture this trend, we generate $\gamma_{\rm NSC}$ from uniform distributions, whose upper bounds depend on $M_{\rm NSC}$. To be specific, we have $\gamma_{\rm NSC}\sim 0.65-2.55$ for $M_{\rm NSC}<M_{\rm crit}$, $\gamma_{\rm NSC}\sim 0.65-1.35$ for $M_{\rm NSC}>10^{8}\ \rm M_{\odot}$, and the upper bound evolves linearly with $\log M_{\rm NSC}$ in-between. Fig.~\ref{rnsc} shows the predictions of our stochastic model of NSC size-mass relations for both $R_{\rm eff}$ (top) and $R_{\rm NSC}$ (bottom). For $R_{\rm eff}$, we show the observational data complied by \citet{neumayer2020nuclear}. For $R_{\rm NSC}$, the results for observed NSCs are derived from Equation~(\ref{e9}) with randomly generated $\gamma_{\rm NSC}$. We further illustrate the size distribution by a mock sample of NSCs with a log-flat mass function in the range of $M_{\rm NSC}\sim 10^{5}-10^{9}\ \rm M_{\odot}$. Similar to the sample of host NSCs produced by our merger trees, for each NSC in the mock sample, both $R_{\rm eff}$ and $\gamma_{\rm NSC}$ are drawn randomly to determine $R_{\rm NSC}$. It turns out that $R_{\rm NSC}$ slowly increases with $M_{\rm NSC}$, and $R_{\rm NSC}\sim 1-20\ \rm pc$ for most NSCs.

Finally, once the NSC properties ($M_{\rm NSC}$, $R_{\rm NSC}$ and $\gamma_{\rm NSC}$) are given, we can calculate the evolution of an in-falling compact object binary. Similar to the galaxy model, the remnant will sink into the NSC centre by DF. We fix the eccentricity $e_{\rm if}$ of the in-fall orbit 0 because the orbit will be mostly circularized by DF in the galaxy when the object reaches the NSC. For simplicity, we assume that NSCs do not evolve with time\footnote{Before compact object binaries merge, their host galaxy may fall into larger structures and even be destroyed/stripped in a galaxy merger. Nevertheless, NSCs can survive such mergers due to their compact nature and appear as compact GCs and ultra compact dwarfs (UCDs). In this work, we assume a 100\% survive rate of NSCs in galaxy mergers for simplicity.}, such that Equation~(\ref{e7}) has an analytical solution:
\begin{align}
    r(t)\simeq r_{0}\left(1-\beta\frac{t}{\tau_{\rm DF,0}}\right)^{1/\beta}\ ,\quad t<\hat{t}_{\rm DF}\equiv \tau_{\rm DF,0}/\beta\ ,\label{e10}
\end{align}
where the sink time $\hat{t}_{\rm DF}$ is the expected time taken for the binary to reach the cluster centre\footnote{A small fraction of highly eccentric binaries can merge before reaching the centre/core.}, $\beta=1.76$ and $\tau_{\rm DF,0}=\tau_{\rm DF}(r_{0}|m_{\bullet},M_{\rm NSC},R_{\rm NSC},\gamma_{\rm NSC},e_{\rm if}=0)$, given the initial distance $r_{0}=3\ \rm pc$. As the binary moves inside the NSC, the stellar environment around it evolves with time according to $r(t)$, with the stellar density $\rho_{\star}(r)$ given by the Dehnen profile (Equ.~\ref{e2}), and the velocity dispersion following \citep{dehnen1993,sedda2020birth}
\begin{align}
    \sigma_{\star}(r)\simeq \sqrt{\frac{GM_{\rm NSC}}{R_{\rm NSC}}}(r/R_{\rm NSC})^{\delta/2}\ ,\quad r\lesssim R\ ,
\end{align}
where $\delta=\gamma_{\rm NSC}$ for $\gamma_{\rm NSC}<1$ and $\delta=2-\gamma_{\rm NSC}$ for $\gamma_{\rm NSC}\ge 1$. Since the Dehnen profile can be divergent at the centre, we assume that each NSC has a core with a radius $r_{\rm c}=R_{\rm NSC}/\Delta_{\rm c}^{1/\gamma_{\rm NSC}}$, defined by the core overdensity parameter $\Delta_{\rm c}$. Once the binary enters the core ($r<r_{\rm c}$), we fix the environment to that evaluated at $r_{\rm c}$. The overdensity parameter $\Delta_{\rm c}$ determines the rate of DH, as most binary evolution happens in the core with high density. We consider two cases with $\Delta_{\rm c}=100$ and 20.

Within certain stellar environments, hard binaries will be further hardened by three-body encounters and GW emission until the final mergers\footnote{For simplicity, we ignore higher-order effects such as exchanges and repeated mergers. Note that in our case the dominant ($\gtrsim 90$\%) hosts of compact object mergers are low-mass NSCs ($M_{\rm NSC}\lesssim 10^{6}\ \rm M_{\odot}$, see Sec.~\ref{s3.1}) which may not be able to retain second-generation BHs efficiently. We also defer the inclusion of central massive BHs to future work (see e.g. \citealt{sedda2020birth}).}, while soft binaries will be disrupted/destroyed. For each binary at galaxy centres, we draw the initial separation $a_{0}$ and eccentricity $e_{0}$ from the catalog of binaries produced by the N-body simulations in \citet{liu2021binary}, matching the total mass $m_{\bullet}$. We only keep track of hard binaries whose initial separations are below the critical separation $a_{\rm HDB}=Gm_{1}m_{2}/\left(m_{\star}\sigma_{\star}^{2}\right)$ \citep{mapelli2021hierarchical}, where we evaluate the stellar environment at $r=R_{\rm NSC}$, since $\sigma_{\star}$ peaks around $R_{\rm NSC}$ and $\rho_{\star}$ drops rapidly at $r\gtrsim R_{\rm NSC}$. We also require $a_{0}<r_{\rm c}$ for efficient DH. The evolution of binary parameters driven by three-body encounters ($|_{\star}$) and emission of GWs ($|_{\rm GW}$) is described by the following differential equations:
\begin{align}
\begin{split}
    \frac{da}{dt}=\left.\frac{da}{dt}\right|_{\rm \star}+\left.\frac{da}{dt}\right|_{\rm GW}\ ,\quad
    \frac{de}{dt}=\left.\frac{de}{dt}\right|_{\star}+\left.\frac{de}{dt}\right|_{\rm GW}\ ,
\end{split}\label{e12}
\end{align}
in which the three-body terms are \citep{sesana2015scattering,mapelli2021hierarchical,sedda2020birth}
\begin{align}
\begin{split}
    \left.\frac{da}{dt}\right|_{\rm \star}=-\frac{GH\rho_{\star}(r)}{\sigma_{\star}(r)}a^{2}\ ,\quad
    \left.\frac{de}{dt}\right|_{\star}=\kappa\frac{GH\rho_{\star}(r)}{\sigma_{\star}(r)}a\ ,
\end{split}\label{e13}
\end{align}
where $H\sim 1-20$ and $\kappa\sim 0.01-0.1$ \citep{sesana2006interaction,sesana2015scattering,mapelli2021hierarchical} are two dimensionless factors. We adopt $H=20$ and explore two cases with $\kappa=0.01$ and 0.1. 
The GW terms can be written as {\citep{peters1964,mapelli2021hierarchical}} 
\begin{align}
\begin{split}
     \left.\frac{da}{dt}\right|_{\rm GW}&=-\frac{64}{5}\frac{A}{a^{3}(1-e^{2})^{7/2}}f_{1}(e)\ ,\\
     \left.\frac{de}{dt}\right|_{\rm GW}&=-\frac{304}{15}e\frac{A}{a^{4}(1-e^{2})^{5/2}}f_{2}(e)\ ,
\end{split}\label{e14}
\end{align}
where $A=G^{3}m_{1}m_{2}(m_{1}+m_{2})/c^{5}$, and
\begin{align}
\begin{split}
    f_{1}(e)=1+\frac{73}{24}e^{2}+\frac{37}{96}e^{4}\ ,\quad
    f_{2}(e)=1+\frac{121}{304}e^{2}\ .
\end{split}
\end{align}

For each binary, we integrate Equations~(\ref{e12}) with the 4th-order Runge-Kutta method for at most 15~Gyr and stop the integration if a merger happens with $a<6G(m_{1}+m_{2})/c^{2}$. Note that in this calculation, we ignore the possibility of ejection by three-body encounters. To take this into account, following \citet{mapelli2021hierarchical}, we compare the separation below which the evolution is dominated by GW emission \citep{mapelli2021hierarchical}
\begin{align}
    a_{\rm GW}=\left[\frac{64}{5H}\frac{A\sigma_{\star}f_{1}(e)}{G\rho_{\star}(1-e^{2})^{7/2}}\right]^{1/5}
\end{align}
with the critical separation below which the binary, if \textit{not} dominated by GW emission, will be ejected by three-body encounters \citep{mapelli2021hierarchical}\footnote{By our definition, the numerical factor $H$ is related to the $\xi$ in the formalism of \citet[see their equ. 11 and 17]{mapelli2021hierarchical} by $H=2\pi\xi$.}
\begin{align}
    a_{\rm ej}=\frac{Hm_{\star}^{2}}{\pi(m_{1}+m_{2})^{3}}\frac{Gm_{1}m_{2}}{v_{\rm esc}^{2}}\ ,
\end{align}
where the escape velocity is given by \citep{fragione2020repeated}
\begin{align}
    \frac{v_{\rm esc}}{\mathrm{km\ s^{-1}}}=40\left(\frac{M_{\rm NSC}}{10^{5}\ \rm M_{\odot}}\right)^{1/3}\left(\frac{\rho_{\star}}{10^{5}\ \rm M_{\odot}\ pc^{-3}}\right)^{1/6}\ .
\end{align}
Here we evaluate $\rho_{\star}$ at $r=R_{\rm NSC}$ as the typical stellar density. If $a_{\rm gw}>a_{\rm ej}$ always holds, as in massive NSCs ($M_{\rm NSC}\gtrsim 10^{6}\ \rm M_{\odot}$), the binary will not be ejected before merger. If $a<a_{\rm ej}<a_{\rm GW}$ occurs, the binary will be ejected at that moment ($t_{\rm ej}$). The subsequent evolution is no longer affected by three-body encounters and the time to merger for such isolated binaries driven by GW emission is 
\begin{align}
    \tau_{\rm GW}= f_{3}(e)\frac{5}{256}\frac{a^{4}(1-e^{2})^{7/2}}{Af_{1}(e)}\ ,\label{e19}
\end{align}
where $f_{3}(e)$ is a factor that captures the circularization of binary orbits by GW emission, and $a=a(t_{\rm ej})$, $e=e(t_{\rm ej})$ for the ejected binary. We adopt $f_{3}(e)=(1-0.8 e)^{-1}$ as an approximation to the result of numerical integration for simplicity\footnote{See equ. 24 in \citet{liu2020gw} for the exact formula for $\tau_{\rm GW}$.}. The delay time $t_{\rm delay}$ is defined as the time between NSC in-fall, $t_{\rm if}$, and the final merger (or end of integration), $t_{\rm mg}$. With this formalism, the population of Pop~III binary remnants in NSCs is mapped to the population of GW events of compact object mergers via $t_{\rm mg}=t_{\rm if}+t_{\rm delay}$. 

\begin{figure}
    \centering
    \includegraphics[width=1\columnwidth]{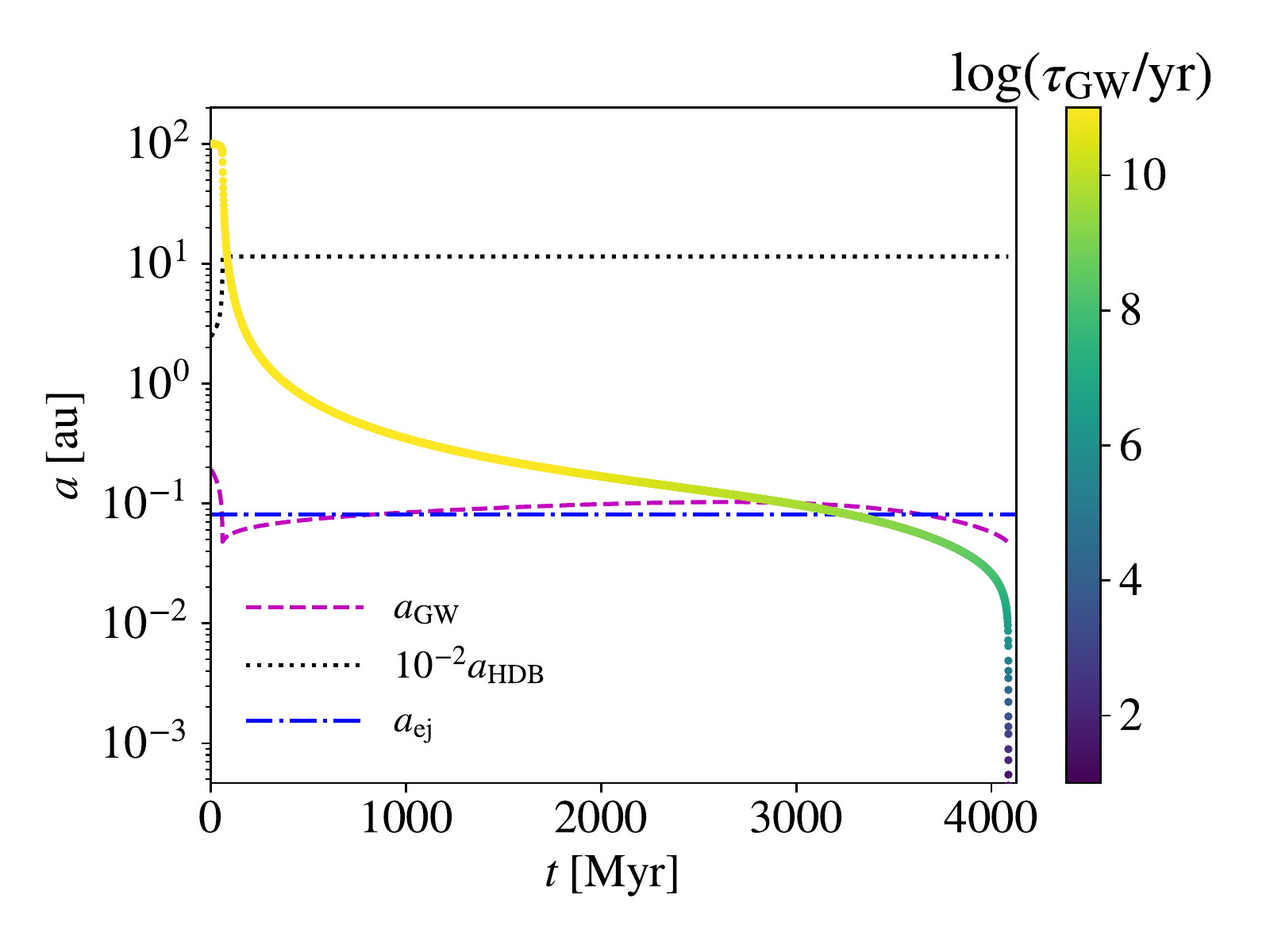}
    \vspace{-25pt}
    \caption{Evolution of binary separation for a NS-BH binary with $m_{1}=20\ \rm M_{\odot}$, $m_{2}=2\ \rm M_{\odot}$, $a_{0}=100\ \rm au$ and $e_{0}=0.1$, in a low-mass NSC with $M_{\rm NSC}=10^{5}\ \rm M_{\odot}$, $R_{\rm NSC}=3\ \rm pc$ and $\gamma_{\rm NSC}=1.5$ (for $\Delta_{\rm c}=100$), assuming no ejection. The $a(t)$ curve is color coded by $\tau_{\rm GW}$. The critical separations for GW-dominance and ejection by three-body encounters are shown with the dashed and dashed-dotted curves, respectively. The critical separation for hard binaries, reduced by a factor of 100, is shown with the dotted curve. The sink time is $\hat{t}_{\rm DF}\simeq 60\ \rm Myr$. According to the criterion, $a<a_{\rm ej}<a_{\rm GW}$, the binary is expected to be ejected at $t\sim 3~\ \rm Gyr$ and merge at $t\sim 4.7\ \rm Gyr$. If not ejected, it will merge at $t\sim 4.1\ \rm Gyr$.}
    \label{at}
\end{figure}

\begin{figure}
    \centering
    \includegraphics[width=1\columnwidth]{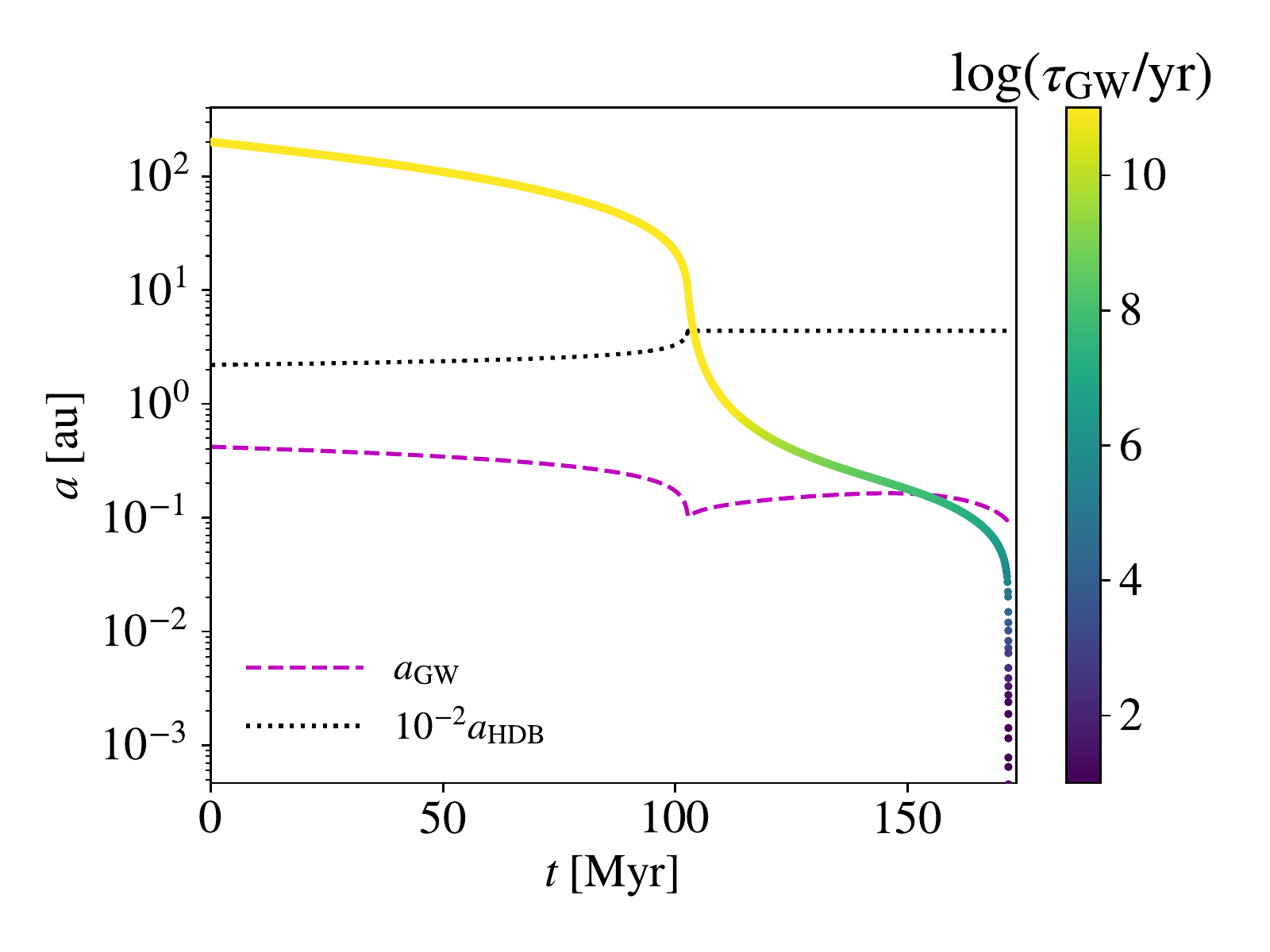}
    \vspace{-25pt}
    \caption{Same as Fig.~\ref{at} but for a BH-BH binary similar to that observed in the event GW190521 ($m_{1}=85\ \rm M_{\odot}$, $m_{2}=66\ \rm M_{\odot}$, $a_{0}=200\ \rm au$, $e_{0}=0$) in a MW-like NSC ($M_{\rm NSC}=2.5\times 10^{7}\ \rm M_{\odot}$, $R_{\rm NSC}=4.4\ \rm pc$, $\gamma_{\rm NSC}=1.8$) with a core density $\rho_{\star}\sim 2.5\times 10^{7}\ \rm M_{\odot}\ pc^{-3}$ at $r\lesssim 0.01\ \rm pc$ \citep{schodel2018distribution}. Here $a_{\rm ej}$ is much smaller than $a_{\rm GW}$ and thus not shown. The sink time is $\hat{t}_{\rm DF}\simeq 100\ \rm Myr$, and the merge time is $t_{\rm mg}\simeq 170\ \rm Myr$.}
    \label{atmw}
\end{figure}

We end this section with two (extreme) examples for the evolution of Pop~III binary remnants in NSCs under $\kappa=0.1$. In the first case, we consider a NS-BH binary with $m_{1}=20\ \rm M_{\odot}$, $m_{2}=2\ \rm M_{\odot}$, $a_{0}=100\ \rm au$ and $e_{0}=0.1$, in a low-mass NSC with $M_{\rm NSC}=10^{5}\ \rm M_{\odot}$, $R_{\rm NSC}=3\ \rm pc$ and $\gamma_{\rm NSC}=1.5$ (for $\Delta_{\rm c}=100$). As shown in Fig.~\ref{at} (assuming no ejection), the binary sinks to the cluster core after $\hat{t}_{\rm DF}\simeq 60\ \rm Myr$, and is ejected at $t_{\rm ej}\sim 3\ \rm Gyr$. The delay time with(out) ejection is $t_{\rm delay}\simeq 4.7\ (4.1)\ \rm Gyr$. 
In the second case, we consider a BH-BH binary similar to that observed in the event GW190521\_030229 (GW190521 for short) with $m_{1}=85\ \rm M_{\odot}$, $m_{2}=66\ \rm M_{\odot}$ \citep{abbott2020gw190521}, $a_{0}=200\ \rm au$ and $e_{0}=0$, in a MW-like NSC ($M_{\rm NSC}=2.5\times 10^{7}\ \rm M_{\odot}$, $R_{\rm NSC}=4.4\ \rm pc$, $\gamma_{\rm NSC}=1.8$) with $\Delta_{\rm c}=1000$. Such a massive binary in a massive dense NSC will not be ejected before merger, for which we have $\hat{t}_{\rm DF}\simeq 100\ \rm Myr$ and $t_{\rm delay}\sim 170\ \rm Myr$, as shown in Fig.~\ref{atmw}. Here $\Delta_{\rm c}=1000$ is chosen to be consistent with observations of the realistic MW NSC by \citet{schodel2018distribution}, which show $\rho_{\star}\sim 2.6\pm 0.3\times 10^{7}\ \rm M_{\odot}\ pc^{-3}$ at $r\lesssim 0.01\ \rm pc$. 


\section{Model predictions}
\label{s3}

In this section, we present the demography, rates and host properties of Pop~III binary remnants falling into NSCs and the subsequent mergers. We explore two sets of parameters for three-body encounters: (i) the optimistic (OP) case for efficient dynamical and GW hardening with $\Delta_{\rm c}=100$, $\kappa=0.1$, and (ii) the conservative (CS) case for moderate hardening with $\Delta_{\rm c}=20$, $\kappa=0.01$. We also consider two assumptions for the occupation frequency of NSCs (see Sec.~\ref{s2.3}): (i) full occupation (F), i.e. $f_{\rm NSC}=1$, in galaxies with $M_{\rm NSC}>10^{5}\ \rm M_{\odot}$, and (ii) partial occupation (P) with $f_{\rm NSC}(M_{\star})<1$, based on local observations (see Fig.~\ref{foccnsc}). Combining these parameter choices leads to 4 NSC models: OP\_F, OP\_P, CS\_F, and CS\_P. 

\begin{figure}
    \centering
    \includegraphics[width=1\columnwidth]{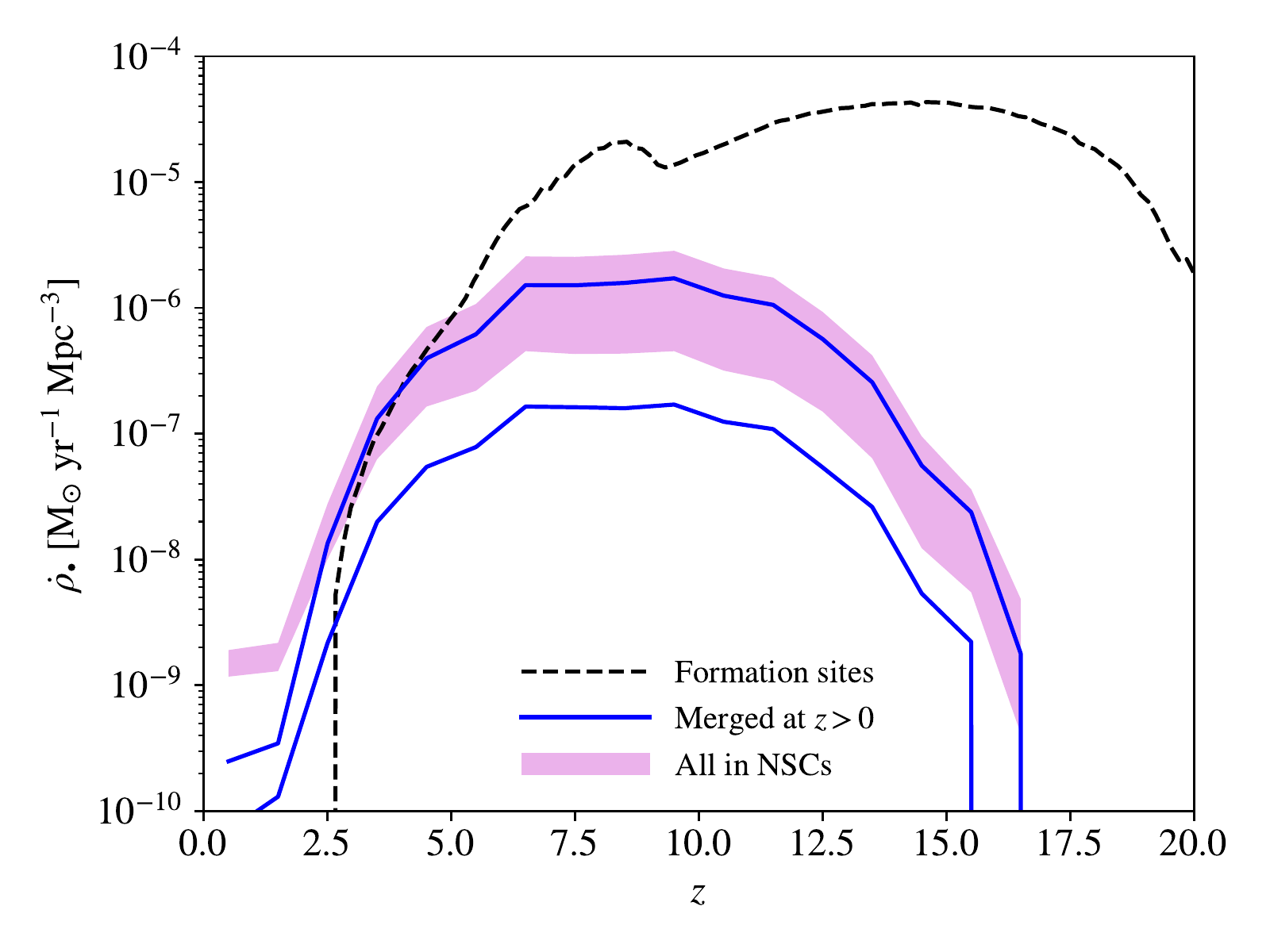}
    \vspace{-25pt}
    \caption{IFRDs of Pop~III binary remnants at galaxy centres ($r<3\ \rm pc$) for the entire sample (shaded region) and those merged at $z>0$ (solid curves). The upper and lower bounds of the former denote the models with full occupation (OP/CS\_F) and observation-based occupation frequency (OP/CS\_P) of NSCs. While for the latter, the upper and lower bounds correspond to OP\_F and CS\_P. To illustrate the delay between remnant formation and in-fall into galaxy centres, we also show the formation rate density of Pop~III binary remnants with the dashed curve, which can be expressed in terms of the SFRD as $\dot{\rho}_{\bullet,\rm form}=f_{\rm B}f_{\rm rem}\rm SFRD_{\rm III}$, where $f_{\rm B}=0.69$ is the mass fraction of binary stars, and $f_{\rm rem}=0.31$ is the overall ratio of remnant mass to initial stellar mass (in binaries).}
    \label{ifrd}
\end{figure}

Fig.~\ref{ifrd} shows the in-fall rate density (IFRD) of Pop~III binary remnants at galaxy centres ($r<3\ \rm pc$) for the entire sample and those merged at $z>0$, on top of the formation rate density of Pop~III binary remnants. It turns out that the in-fall of Pop~III binary remnants into galaxy centres starts at $z\sim 15-17$, and the IFRD peaks around $z\sim 6-10$, with a shape that is almost independent of NSC models. Note that most Pop~III SF (as well as remnant formation) happens at $z\sim 6-20$ with 2 peaks at $z\sim 8$ and 14. The delay between remnant formation and in-fall reflects the time taken for (Pop~I/II) galaxies/NSCs to form and grow, and DF to drive the in-spiral of remnants. 

In Sec.~\ref{s3.1}, we discuss the host galaxy and NSC properties for the entire sample of Pop~III binary remnants at galaxy centres as well as those merged at $z>0$. In Sec.~\ref{s3.2}, we focus on the mergers of Pop~III remnants in NSCs, predicting their statistics and cosmic MRD.

\subsection{Host systems of Pop~III remnants}
\label{s3.1}

\begin{figure}
    \centering
    \includegraphics[width=1\columnwidth]{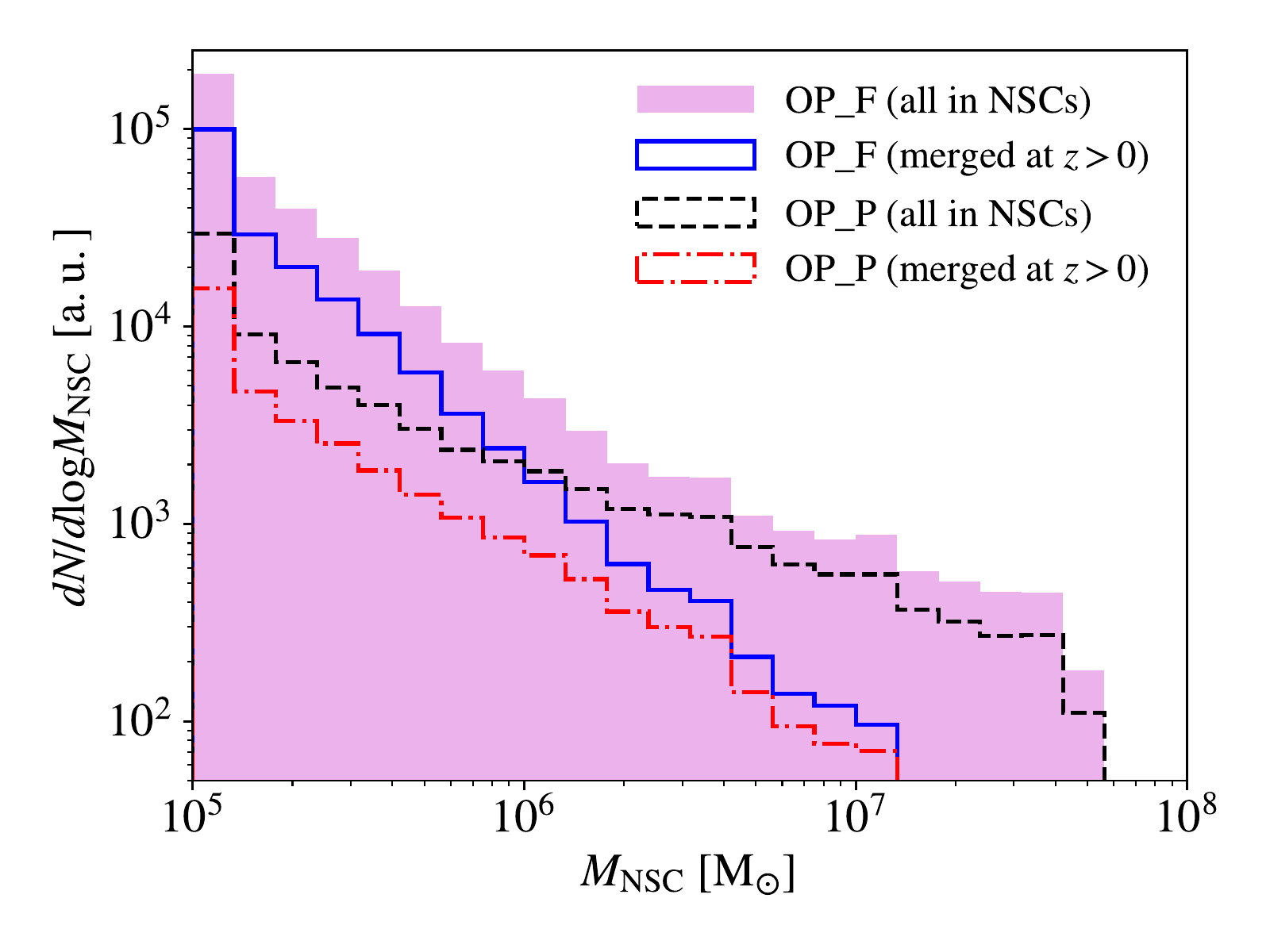}
    \vspace{-25pt}
    \caption{Distributions of NSC masses hosting Pop~III binary remnants for the OP models, for the entire sample of NSCs and those merged at $z>0$, with full occupation (histograms and the solid contour) and observation-based occupation frequency (dashed and dashed-dotted contours). It is evident that low-mass NSCs are the dominant hosts, such that $90$\% of remnants merged at $z>0$ are from NSCs with $M_{\rm NSC}\lesssim 3.9\ (8.1)\times 10^{5}\ \rm M_{\odot}$ in the OP\_F (P) model.}
    \label{mnscdis}
\end{figure}

\begin{figure}
    \centering
    \includegraphics[width=1\columnwidth]{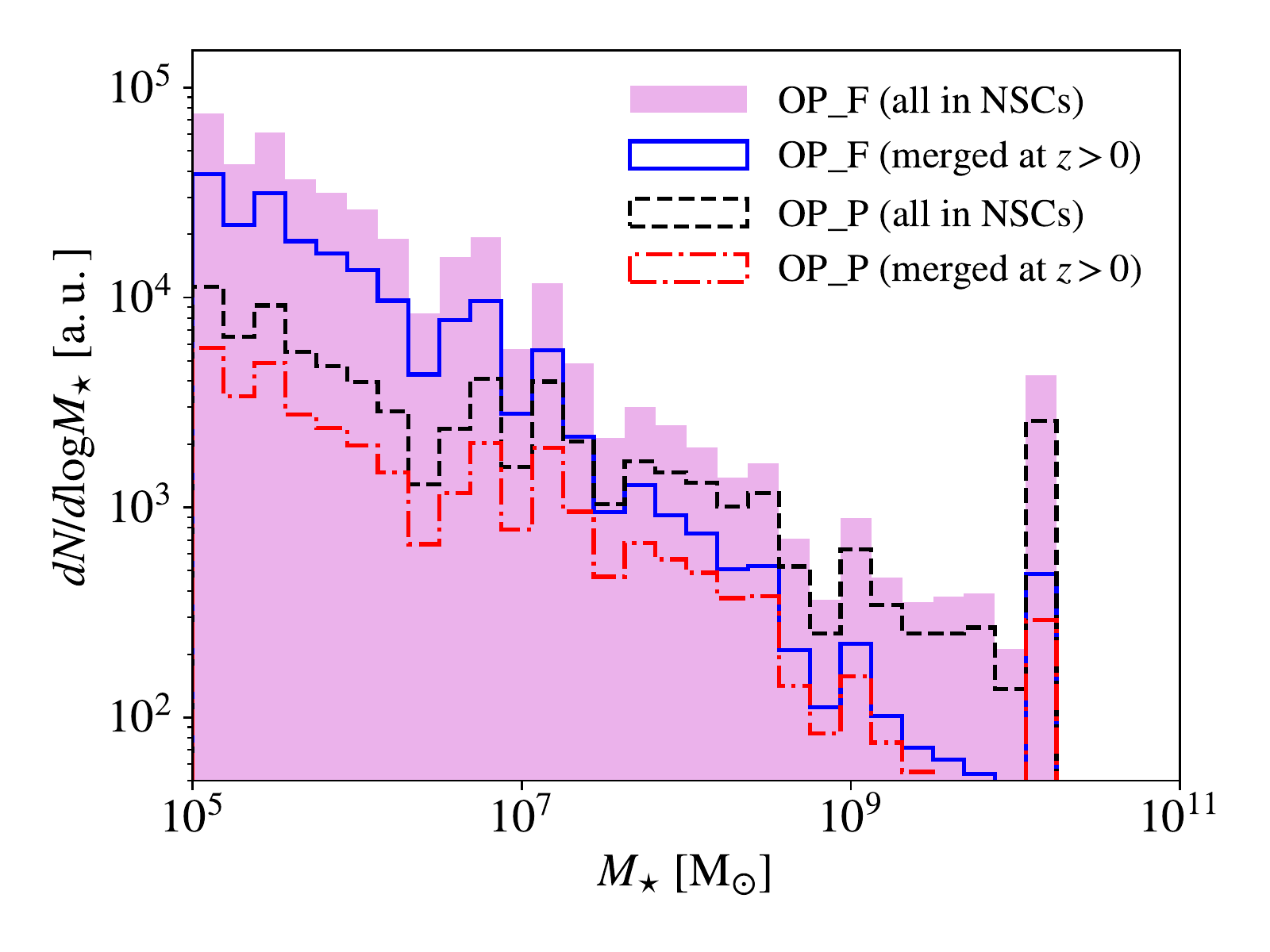}
    \vspace{-25pt}
    \caption{Distributions of galaxy (stellar) masses containing NSCs that host Pop~III binary remnants from the OP models, for the entire sample and those merged at $z>0$ with full occupation (histograms and the solid contour) and observation-based occupation frequency (dashed and dashed-dotted contours). As can be seen, Pop~III remnants mostly fall into the centres of low-mass dwarf galaxies, such that $90$\% of remnants merged at $z>0$ are from galaxies with $M_{\star}\lesssim 7.5\ (34)\times 10^{6}\ \rm M_{\odot}$ in the OP\_F (P) model.}
    \label{mgdis}
\end{figure}

\begin{figure}
    \centering
    \includegraphics[width=1\columnwidth]{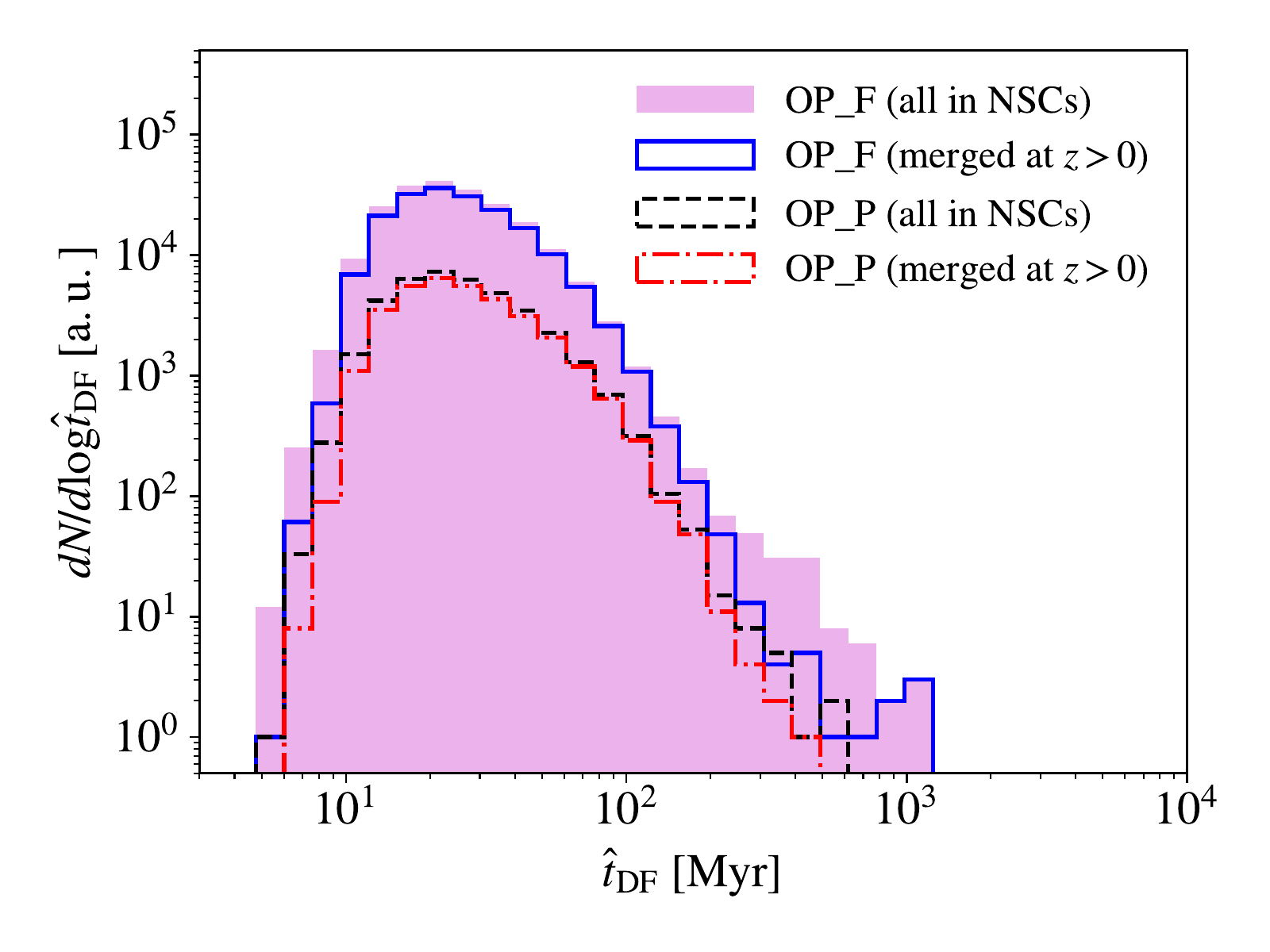}
    \vspace{-25pt}
    \caption{Sink time distributions of Pop~III binary remnants in NSCs from the OP models, for the entire sample in NSCs and those merged at $z>0$ with full occupation (histograms and the solid contour) and observation-based occupation frequency (dashed and dashed-dotted contours) of NSCs.}
    \label{tdfdis}
\end{figure}

For conciseness, here we only show the results from the OP models for different NSC occupation assumptions, as the trends seen in the CS models are similar. 
Fig.~\ref{mnscdis} and \ref{mgdis} show the distributions (weighted by the number of binary remnants) of NSC masses hosting Pop~III binary remnants, and of the corresponding galaxy stellar masses (at the moment when the binary reaches the centre), respectively. In all cases, the distributions are significantly biased towards low-mass systems, especially for the binaries merged at $z>0$. The reason is that DF is more efficient in low-mass galaxies (see Fig.~\ref{tdf}), and Pop~III binary remnants are more likely to be destroyed in more massive NSCs with higher $\sigma_{\star}$. Actually, assuming full occupation of NSCs, $50\ (90)\%$ of Pop~III binary remnant \textit{mergers} are hosted by NSCs with $M_{\rm NSC}\lesssim 1.2\ (3.9)\times 10^{5}\ \rm M_{\odot}$, (initially) in galaxies with $M_{\star}\lesssim 4.5\ (75.0)\times 10^{5}\ \rm M_{\odot}$. With the observation-based occupation frequency, the mass distributions are less dominated by small systems, such that we have $M_{\rm NSC}\lesssim 1.4\ (8.1)\times 10^{5}\ \rm M_{\odot}$ and $M_{\star}\lesssim 7.0\ (340)\times 10^{5}\ \rm M_{\odot}$ for $50\ (90)\%$ of mergers. 

Note that the dominant (initial) host galaxies of our Pop~III binary remnant mergers are high-redshift ($z\gtrsim 4$) dwarf galaxies ($M_{\star}\lesssim 10^{7}\ \rm M_{\odot}$) which will further grow and merge into larger galaxies. Some of them will even be destroyed/stripped in galaxy mergers where the NSCs can survive as compact GCs and ultra compact dwarfs (UCDs). As the final merger can be significantly delayed with respect to the in-fall (see below), we expect a more diverse population of host systems at the moments of merger, including NSCs of more massive central galaxies and satellites, GCs and UCDs, compared with the initial hosts at the moments of in-fall. We defer a detailed investigation into the host systems at the moments of merger to future work. 

Fig.~\ref{tdfdis} shows the distributions of sink time $\hat{t}_{\rm DF}$ (see Equ.~\ref{e10} for definition) of Pop~III binary remnants in NSCs. In general, we have $\hat{t}_{\rm DF}\sim 5-10^{3}\ \rm Myr$, and the distribution peaks at $\hat{t}_{\rm DF}\sim 20\ \rm Myr$. The results from different NSC occupation models are similar, although the distribution is narrower in the observation-based model (with $\hat{t}_{\rm DF}\lesssim 500\ \rm Myr$) than that with full occupation. When restricted to the binaries merged at $z>0$, cases with $\hat{t}_{\rm DF}\gtrsim 200\ \rm Myr$ are reduced, especially for the full occupation model. We expect these cases to be extreme low-mass, low-density NSCs where three-body hardening is inefficient. This reduction is more efficient in the CS models.

\subsection{Mergers of Pop~III remnants in NSCs}
\label{s3.2}

\begin{table}
    \caption{Key characteristics of Pop~III binary remnant mergers for 4 NSC models, where $\Delta_{\rm c}$ is the core overdensity parameter that measures how dense the NSC core is compared with the average, $\kappa$ is the dimensionless efficiency at which DH increases the binary eccentricity (see Equ.~\ref{e13}), $f_{\rm NSC}$ is the NSC occupation frequency (see Fig.~\ref{foccnsc} for the case of $f_{\rm NSC}<1$), and $f_{\rm mg}$ is the fraction of Pop~III binary remnants in galaxy centres ($r<3\ \rm pc$) that eventually merge at $z>0$. The peak MRD and the corresponding peak redshift $z_{\rm peak}$ are also shown (see Fig.~\ref{mrdcomp} for details). Note that on average $760$ Pop~III binary remnants will fall into galaxy centres out of the $2.1\times 10^{5}\ \rm M_{\odot}$ of Pop~III stars formed in a merger tree, the efficiency of GW events $\epsilon_{\rm GW}$, defined as the number of compact object mergers (at $z>0$) per unit stellar mass, can be related to the merge fraction $f_{\rm mg}$ with $\epsilon_{\rm GW}\simeq 0.0036f_{\rm mg}\ \rm M_{\odot}^{-1}$.}
    \centering
    \begin{tabular}{ccccccc}
        \hline
        Model & $\Delta_{\rm c}$ & $\kappa$ & $f_{\rm NSC}$ & $f_{\rm mg}$ & Peak MRD & $z_{\rm peak}$ \\
        & & & & & $[\rm yr^{-1}\ Gpc^{-3}]$\\
        \hline
        OP\_F & 100 & 0.1 & 1 & 0.50 & 10.4 & 7.2 \\
        OP\_P & 100 & 0.1 & $<1$ & 0.090 & 1.78 & 7.2\\
        CS\_F & 20 & 0.01 & 1 & 0.30 & 2.28 & 5.2\\
        CS\_P & 20 & 0.01 & $<1$ & 0.055 & 0.378 & 5.2\\
        \hline
    \end{tabular}
    \label{t1}
\end{table}

In Table~\ref{t1}, we summarize the key characteristics of Pop~III binary remnant mergers (in NSCs) for different NSC models. As mentioned in Sec.~\ref{s2.2}, $7.5\%$ of Pop~III remnants will fall into the centres ($r<3\ \rm pc$) of galaxies with $M_{\rm NSC}>10^{5}\ \rm M_{\odot}$, of which $\sim  5.5-50\%$ can merge at $z>0$. The corresponding efficiency of GW events $\epsilon_{\rm GW}$, defined as the number of compact object mergers (at $z>0$) per unit stellar mass, takes the range of $0.2-2\times 10^{-4}\ \rm M_{\odot}$, which is comparable to the values in the classical binary stellar evolution (BSE) channel \citep{kinugawa2014possible,hartwig2016,belczynski2017likelihood,hijikawa2021population}. We also find that ejection of binaries by three-body encounters is rare ($\sim 1\%$), independent of NSC models. 

\subsubsection{Delay time and binary statistics}
\label{s3.2.1}

\begin{figure}
    \centering
    \includegraphics[width=1\columnwidth]{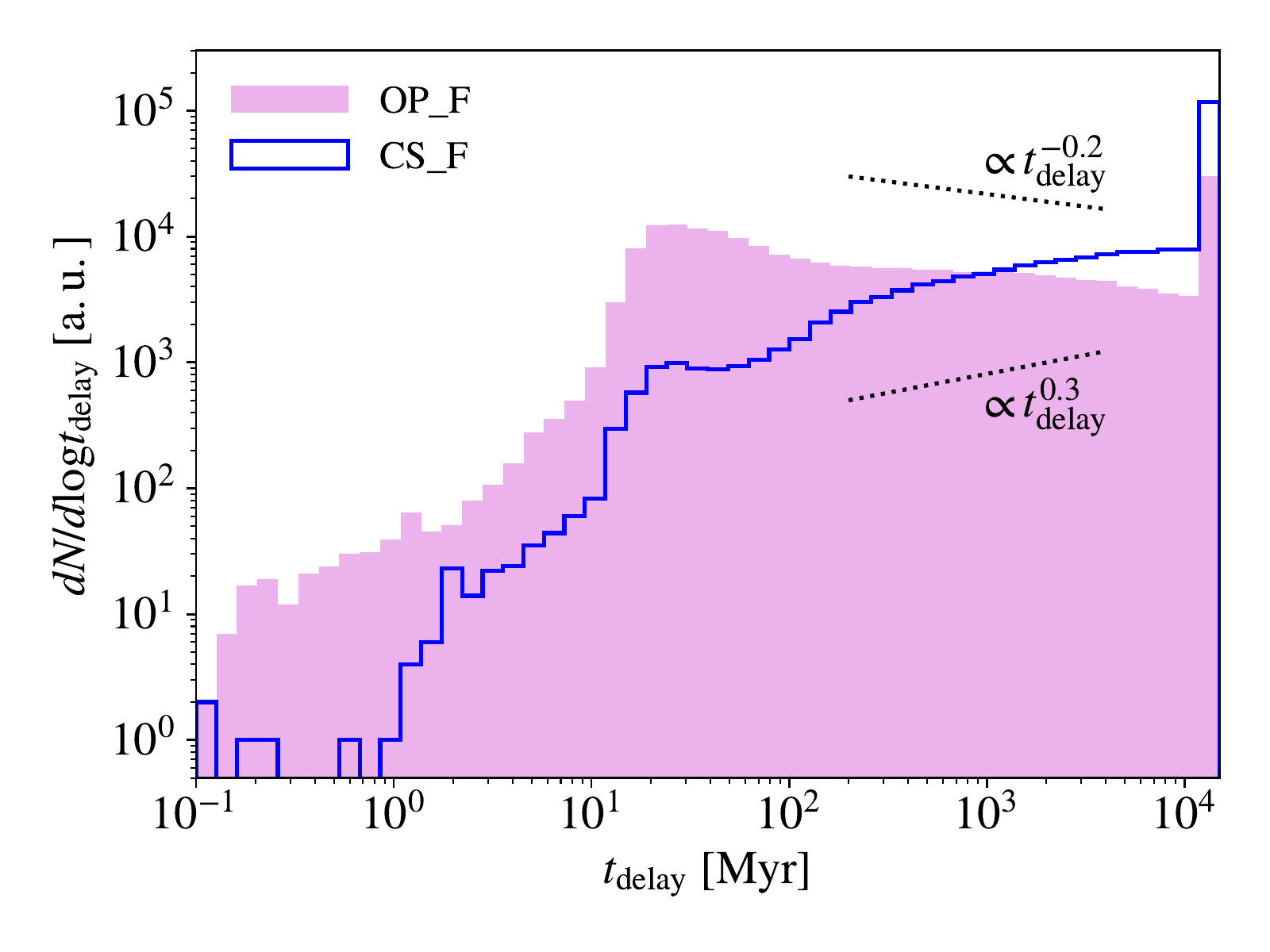}
    \vspace{-25pt}
    \caption{Delay time distributions of Pop~III \textit{hard} binary remnants in NSCs for the OP (histograms) and CS (solid contour) scenarios, with full occupation of NSCs. Power-law approximations for the distributions at $t_{\rm delay}\gtrsim 20\ \rm Myr$ are shown with the dotted lines. The tail at $t<20\ \rm Myr$ makes up $\simeq 7.6\ (0.6)\%$ of all mergers in the OP (CS)\_F model. Here, the binaries with $t_{\rm delay}>15\ \rm Gyr$ are pilled up at the last bin of $t_{\rm delay}\sim 15\ \rm Gyr$. }
    \label{tdelay}
\end{figure}

\begin{figure}
    \centering
    \includegraphics[width=1\columnwidth]{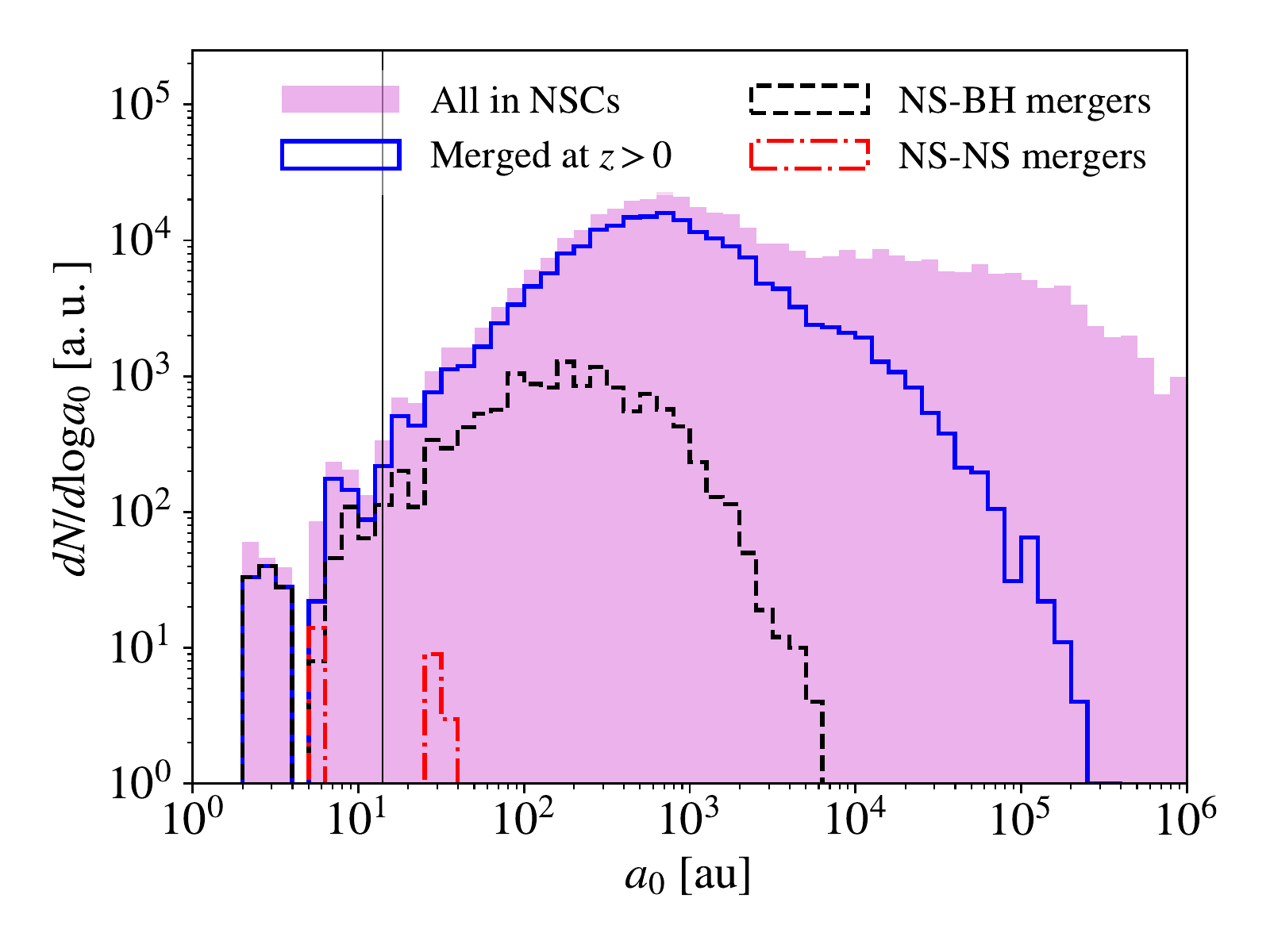}
    \vspace{-25pt}
    \caption{Distributions of initial separation of Pop~III binary remnants in NSCs from the OP\_F model, for the entire sample (histograms) and those merged at $z>0$ (solid contour). The sub-samples of NS-BH and NS-NS mergers are also shown with the dashed and dashed-dotted contours, accounting for 6.7\% and 0.014\% of all mergers, respectively. The thin vertical line denotes the typical maximum radius of massive Pop~III stars, $3000\ \rm R_{\odot}$, which is approximately the critical separation for close binary interactions (such as common envelope evolution).}
    \label{adis}
\end{figure}

\begin{figure}
    \centering
    \includegraphics[width=1\columnwidth]{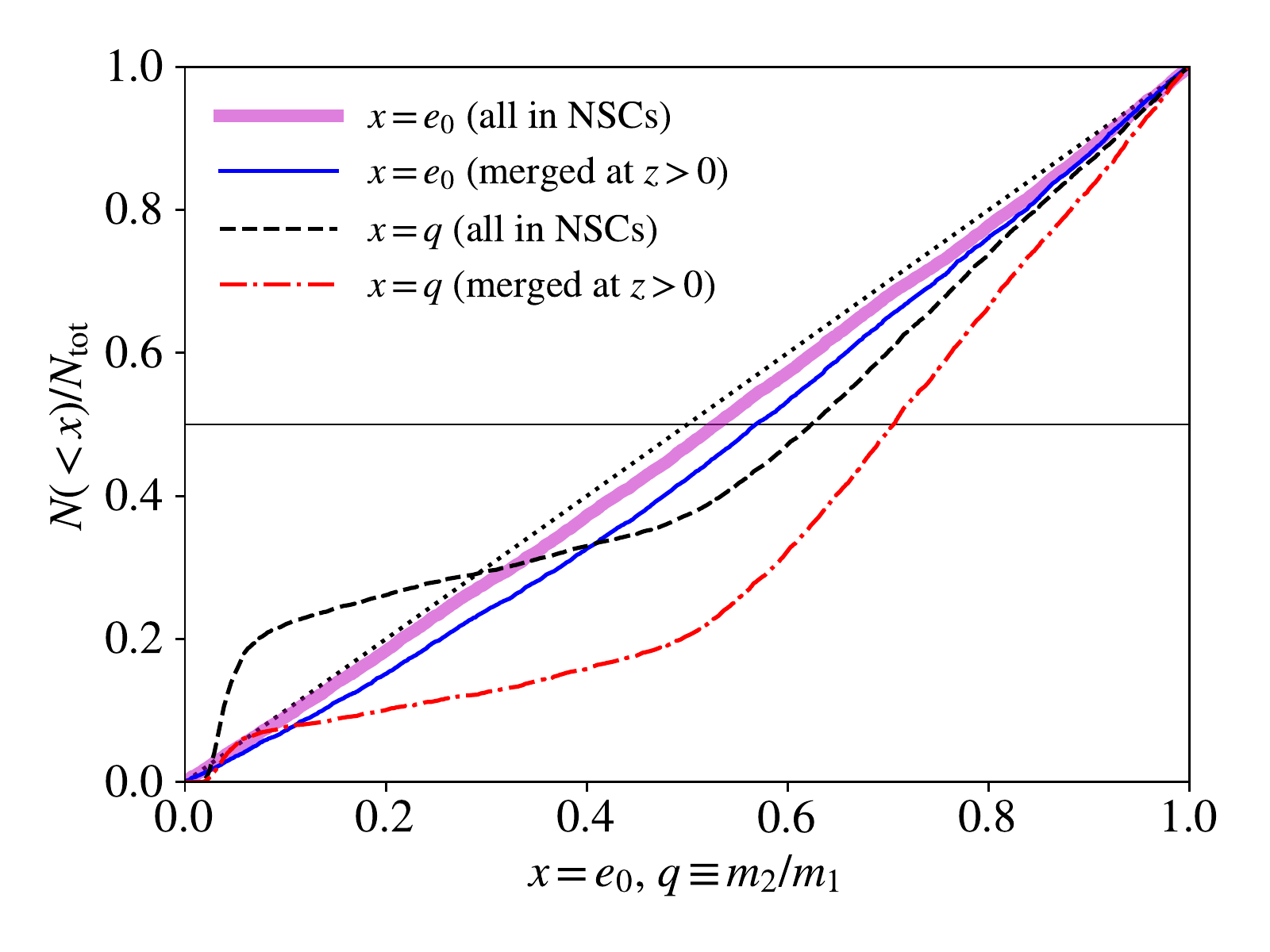}
    \vspace{-25pt}
    \caption{Cumulative distributions of initial eccentricity and mass ratio of Pop~III binary remnants in NSCs from the OP\_F model, for the entire sample (thick solid and dashed) and those merged at $z>0$ (solid and dashed-dotted). The dotted line denotes the case of a uniform distribution in [0,1]. The thin horizontal line is plotted to point out the medians. The mass ratio distribution for binaries merged at $z>0$ is more biased to the high end than that for the entire sample since low-$q$ binaries are less massive and thus merge less efficiently in NSCs.}
    \label{qedis}
\end{figure}

Now we consider the distributions of delay times\footnote{In our case, the delay time is defined as the time between \textit{NSC in-fall} and the final merger, instead of the time between the initial remnant/star formation and merger. Therefore, the MRD is a convolution of the delay time distribution with the IFRD (see Fig.~\ref{ifrd}) rather than the SFRD.} and binary orbital parameters for our Pop~III binary remnants. Fig.~\ref{tdelay} shows the delay time distributions of \textit{hard}\footnote{On average 60\% of our Pop~III binary remnants in NSCs are hard binaries.} Pop~III binary remnants for the OP and CS models with full occupation of NSCs. The results with the observation-based occupation frequency of NSCs are similar and not shown. Among these hard binaries, $\sim 90\ (50)\%$ can merge at $z>0$ under the OP (CS) hardening. In both cases, the delay time distribution is almost log-flat at $t_{\rm delay}\gtrsim 20\ \rm Myr$, while $t_{\rm delay}$ is generally shorter in OP models than in CS models. When approximated with a power law, $dN
/d\log t_{\rm delay}\propto t_{\rm delay}^{\zeta}$, the distribution has a power-law slope $\zeta\sim -0.2$ with the OP hardening and $\zeta\sim 0.3$ for the CS case. Moreover, there is a tail of lower delay times extending down to $t_{\rm delay}\sim 0.1\ \rm Myr$, which is more important with more efficient hardening. For instance, the fraction of hard binaries with $t_{\rm delay}<20\ \rm Myr$ is $\simeq 0.076$ (0.006) for the OP (CS) models. We expect this tail to be made of binaries with high initial eccentricities ($e_{0}\gtrsim 0.9$) that can merge even before reaching the NSC core due to hardening by GW emission (see Equs.~\ref{e14}). 

For the statistics of binary orbital parameters of our Pop~III binary remnants in NSCs, we only show the results of the OP\_F model since similar trends are found in other NSC models. Fig.~\ref{adis} shows the distributions of initial separation $a_{0}$ of Pop~III binary remnants in NSCs (including sub-samples of NS-NS\footnote{We assume that remnants with masses less than $3\ \rm M_{\odot}$ are NSs.} and NS-BH mergers). For the entire sample in NSCs, the distribution is very broad, ranging from a few au to a few pc, which is dominated ($\gtrsim 99.7\%$) by wide binaries ($a_{0}\gtrsim 10\ \rm au$)\footnote{As discussed in \citet{liu2021binary}, Pop~III star clusters expand during the protostellar accretion phase due to angular momentum conservation, making it difficult to form close binaries of Pop~III stars/remnants.}. For such wide-binaries, close binary interactions are likely inactive such that they cannot merge within a Hubble time if isolated. However, with DH in NSCs, binaries with initial separations up to $2\times 10^{5}\ \rm au$ can still merge at $z>0$, making up about half of the entire sample in NSCs. These mergers are dominated by BH-BH binaries ($\gtrsim 90\%$), while NS-BH and NS-NS mergers only count for $\sim 4-7\%$ and $\lesssim 0.014\%$ of all GW events. NS-BH and NS-NS mergers also have lower initial separations ($a_{0}\lesssim 6000$ and 40~au, respectively), compared with the dominant BH-BH mergers. This is due to the fact that in young Pop~III star clusters, low-mass binaries tend to be disrupted by encounters with massive stars, leading to lower initial fractions of binaries involving NSs. Besides, such binaries with relatively lower masses can be more easily destroyed in massive NSCs, and the hardening from GW emission is also less efficient. 

Fig.~\ref{qedis} displays the cumulative distributions of initial eccentricity $e_{0}$ and mass ratio $q$ of Pop~III binary remnants in NSCs. The distribution of $e_{0}$ is almost uniform for the entire sample, consistent with the results from N-body simulations \citep{liu2021binary}. For binaries merged at $z>0$, on the other hand, the distribution is slightly biased towards higher $e_{0}$ since hardening by GW emission is more efficient for eccentric binaries (see Equs.~\ref{e14} and \ref{e19}). This bias is slightly more pronounced for the CS models, such that the median initial eccentricity of mergers is 0.57 (0.60) with OP (CS) hardening. Interestingly, we find that the initial eccentricity does not correlate with $a_{0}$ and $q$. 

The mass ratio distribution is more complex, with three groups of binaries at $q\sim$ (a) $0.02-0.07$, (b) $0.07-0.5$ and (c) $0.5-1$. Each regime can be approximated with a uniform distribution. For Group (a) binaries, the primaries are mostly direct-collapse BHs with $m_{1}\gtrsim 25\ \rm M_{\odot}$, and the secondaries are NSs with $m_{2}\sim 1.4-3\ \rm M_{\odot}$. In Group (b), most binaries are made of two BHs, one with $m_{2}\sim 5-40\ \rm M_{\odot}$ including those (with $m_{2}\sim 5- 25\ \rm M_{\odot}$) born in core-collapse supernovae, and another with $m_{1}\sim 40-85\ \rm M_{\odot}$, including those (with $m_{1}\sim 40-57\ \rm M_{\odot}$) from PPISNe. Finally, Group (c) binaries are mostly made of massive BHs with $m_{1}\sim m_{2}\sim 40-85\ \rm M_{\odot}$. 
When restricted to the binaries merged at $z>0$, the median mass ratio shifts from $0.6$ (for the entire sample in NSCs) to 0.7. The latter is actually consistent with the median mass ratio of GW events detected by the LIGO collaboration \citep{abbott2020population,kinugawa2021}. Meanwhile, the fraction of Group (a) binaries is significantly reduced (from $\sim 20\%$ to $\lesssim 10\%$), consistent with the trend seen in the distribution of $a_{0}$, since it is more difficult for lower-mass binaries to merge in NSCs. 

\subsubsection{Comparison with the BSE channel and LIGO events}
\label{s3.2.2}

\begin{figure}
    \centering
    \includegraphics[width=1\columnwidth]{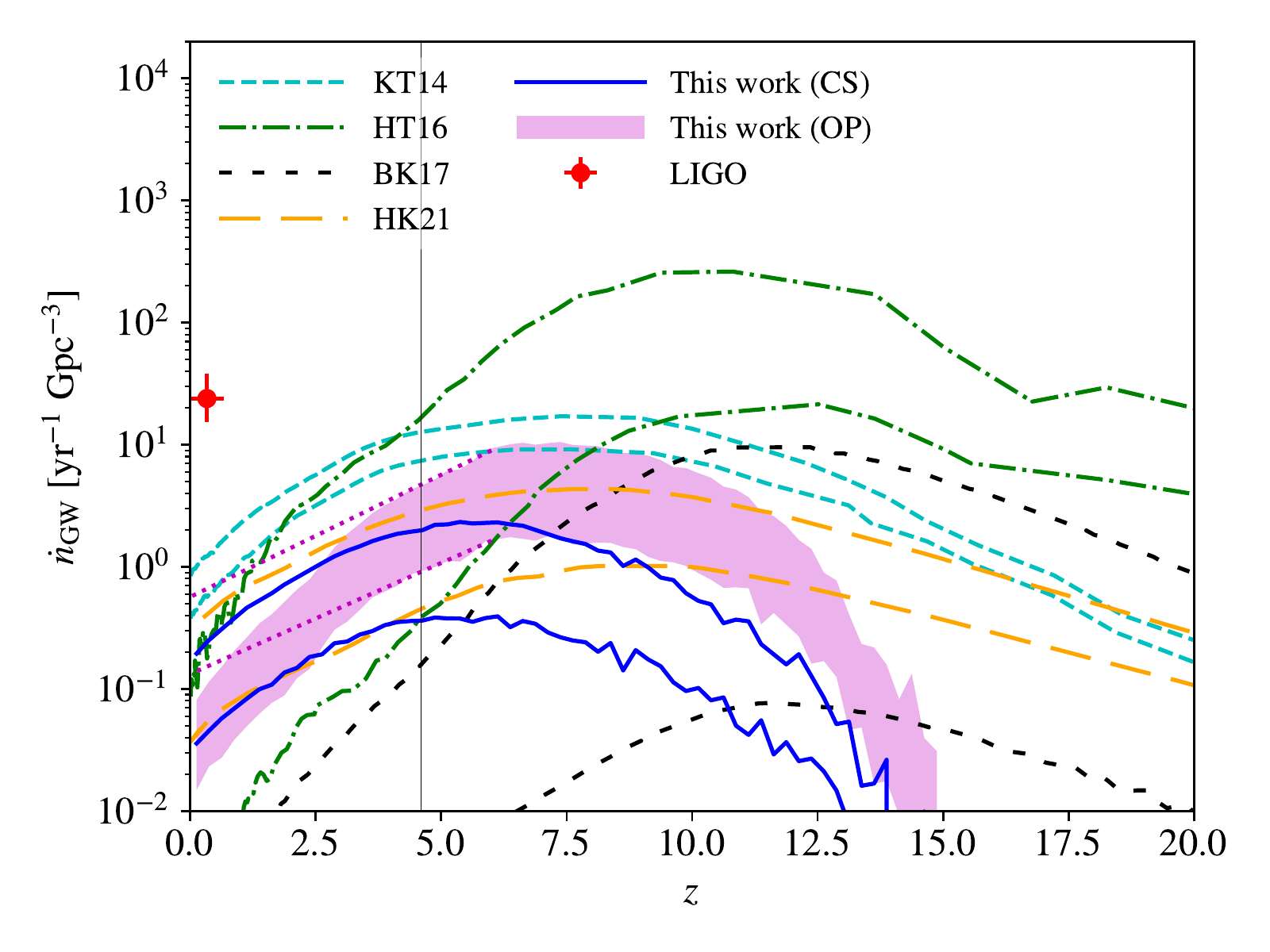}
    \vspace{-25pt}
    \caption{MRDs of Pop~III binary remnants in NSCs for our OP (shaded region) and CS (solid) models with upper and lower bounds corresponding to full occupation and observation-based frequency of NSCs. The extrapolations of the MRDs from the OP models are shown with the dotted lines. For comparison, we show the results of 4 studies for the BSE channel, which are scaled according to our $\rm ISMD_{\rm III}$: \citet[KT14, dashed]{kinugawa2014possible}, \citet[HT16, dashed-dotted]{hartwig2016}, \citet[BK17, sparsely-dashed]{belczynski2017likelihood} and \citet[HK21, long-dashed]{hijikawa2021population}. The local rate density inferred from the LIGO events, $\dot{n}_{\rm GW}(z\sim 0.33)=23.9_{-8.6}^{+14.3}\ \rm yr^{-1}\ Gpc^{-3}$ \citep{abbott2020population}, is shown with the (red-crossed) symbol (for 90\% confidence interval). The thin vertical line marks the turn-over redshift of the target halo, $z\simeq 4.6$, below which our results may not be cosmologically representative.}
    \label{mrdcomp}
\end{figure}

Next, we compare our predictions for Pop~III binary remnant mergers in NSCs with the results for the BSE channel \citep{kinugawa2014possible,hartwig2016,belczynski2017likelihood,hijikawa2021population} and observations by LIGO \citep{abbott2020population,abbott2020gw190521,3ogc}. In Fig.~\ref{mrdcomp}, we show the MRD redshift evolution for our NSC models. Since our merger trees are only representative at $z\gtrsim 5$, the low-$z$ rate can be underestimated, especially for the OP models with shorter delay times. We extrapolate the MRDs from $z\sim 5$ to $z=0$ in the OP models, to approximate the evolution in the low-$z$ regime. For each work considered for comparison \citep{kinugawa2014possible,hartwig2016,belczynski2017likelihood,hijikawa2021population}, we plot the lower and upper bounds for their different assumptions. As mentioned in Sec.~\ref{s2.1.1}, we here constrain the Pop~III SF history by scaling the SFRDs (and MRDs) of these studies such that the scaled integrated Pop~III stellar mass densities ($\mathrm{ISMD}_{\rm III}\equiv\int_{0}^{\infty}\mathrm{SFRD}_{\rm III}|dt/dz|dz$) are identical to our value $\simeq 7\times 10^{4}\ \mathrm{M_{\odot}\ Mpc^{-3}}$. In general, the MRD of Pop~III binary remnants in NSCs is comparable to those for the BSE channel, especially in the post-reionization era ($z\lesssim 6$). For our DH channel with NSCs, the local ($z\sim 0$) MRD is $\sim 0.02-0.6\ \rm yr^{-1}\ Gpc^{-3}$, within the range $\sim 0.01-1\ \rm yr^{-1}\ Gpc^{-3}$ predicted by most BSE models. However, our MRD only starts at $z\sim 15$ and peaks at a later time ($z\sim 6$) compared with most BSE MRDs that extend to very high redshifts ($z\sim 20$) with non-negligible rates ($\gtrsim 0.01\ \rm yr^{-1}\ Gpc^{-3}$). The reason is that galaxy/NSC formation/growth and in-spiral of remnants by DF take time. Measuring the MRD to very high redshifts ($z\gtrsim 15$) can reveal the relative importance of the two channels. 

In all the models for Pop~III mergers considered here (see Fig.~\ref{mrdcomp}), with the Pop~III SF history constrained by the \textit{Planck} measurement of the Thomson optical depth, the local MRD is lower than that of LIGO events, $\dot{n}_{\rm GW}(z\sim 0.33)=23.9_{-8.6}^{+14.3}\ \rm yr^{-1}\ Gpc^{-3}$ \citep{abbott2020population}, by at least one order of magnitude. Given our predicted local MRD $\sim 0.02-0.6\ \rm yr^{-1}\ Gpc^{-3}$, up to $\sim 4\%$ (2) of the 57 events detected by LIGO so far are expected to originate from Pop~III binary remnants in NSCs. Note that our merger trees are only cosmologically representative at $z\gtrsim 5$, such that the local MRD can be underestimated since we do not fully take into account low-mass ($M_{\rm h}\lesssim 10^{12}\ \rm M_{\odot}$) haloes at lower redshifts, especially for $z<2.5$. The actual fraction of LIGO events from Pop~III binary remnants in NSCs could thus be higher. Besides, most of our predicted GW events from Pop~III binary remnant mergers in NSCs can be detected at promising rates by the 3rd generation of GW instruments capable of reaching $z\gtrsim 10$, planned for the next decades, such as the Einstein Telescope \citep{hild2009xylophone} and Decihertz Observatory \citep{dechihertz}. For instance, we predict a full-sky detection rate of $\sim 170-2720\ \rm yr^{-1}$ for a horizon redshift $z_{\max}=10$. 

\begin{figure}
    \centering
    \includegraphics[width=1\columnwidth]{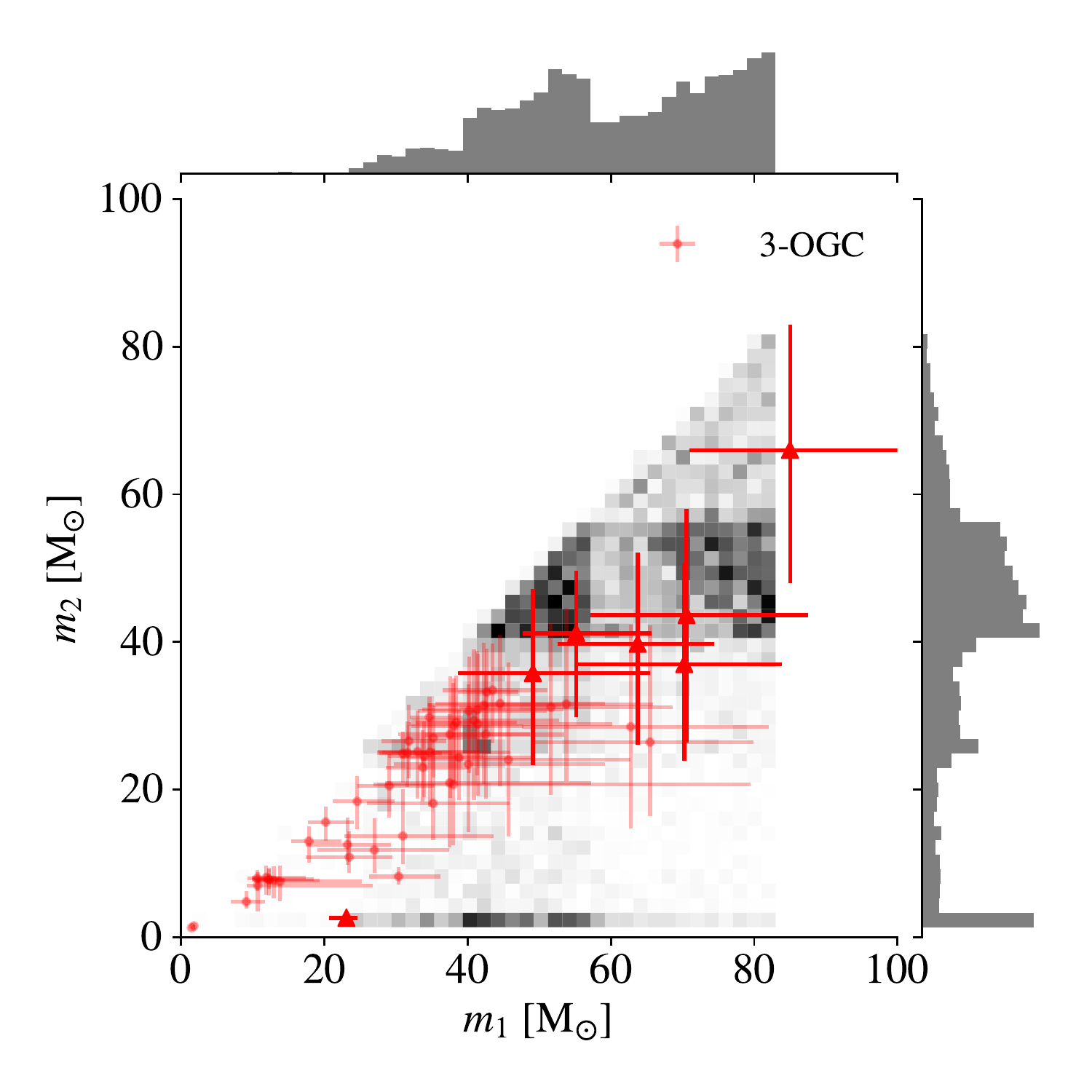}
    \includegraphics[width=1\columnwidth]{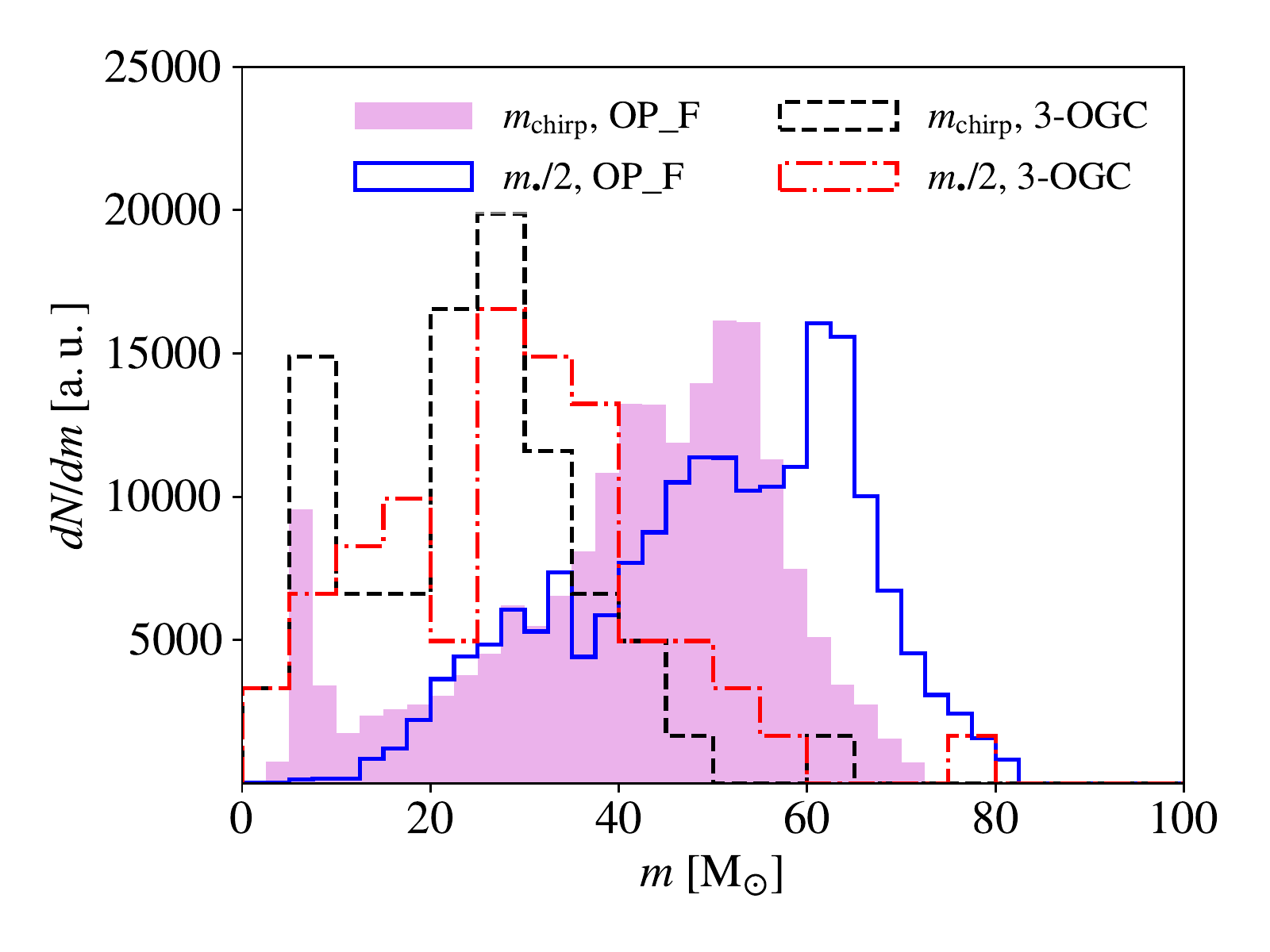}
    \vspace{-25pt}
    \caption{Mass distributions of Pop~III binary remnant mergers in NSCs from the OP\_F model. \textit{Top panel}: joint distribution of primary and secondary masses, where the 57 events from the 3-OGC \citep{3ogc} are plotted as data points with 90\% confidence intervals. We highlight with triangles the events involving massive BHs that are most likely to have Pop~III progenitors, as well as the special event GW190814 that may be a NS-BH merger (see the main text for details). \textit{Bottom pandel}: distributions of chirp mass (histograms) and total mass (solid contour), where the corresponding LIGO observations \citep{3ogc} are shown with the dashed and dashed-dotted contours, respectively. All distributions are normalized by the total number of events. Here we use the results from the NRSur7dq4 waveform model for the event GW190521 \citep{abbott2020gw190521}. Note that the horizontal axis for the total mass is $m_{\bullet}/2$.}
    \label{mdis}
\end{figure}

Fig.~\ref{mdis} shows the joint distribution of primary and secondary masses $m_{1}$ and $m_{2}$ (top) as well as distributions of chirp mass $m_{\rm chirp}\equiv (m_{1}m_{2})^{3/5}/(m_{1}+m_{2})^{1/5}$ and total mass $m_{\bullet}\equiv m_{1}+m_{2}$ (bottom) for the Pop~III binary remnant mergers in the OP\_F model, where the statistics of the 57 events from the the third Open Gravitational-wave Catalog \citep[3-OGC]{3ogc} are also shown for comparison. We again use OP\_F as the example since other models predict similar distributions. Here, for the special event GW190521 that may involve two BHs in the PPISN mass gap for Pop~I/II stars, we show the results from the NRSur7dq4 waveform model with $m_{1}=85_{-14}^{+21}\rm M_{\odot}$ and $m_{2}=66_{-18}^{+17}\ \rm M_{\odot}$ \citep{abbott2020gw190521}. Most of the mergers in our models are made of massive BHs with $m_{1}\sim m_{2}\sim 40-85\ \rm M_{\odot}$. Particularly, two regions in the $m_{2}$-$m_{1}$
space are most densely populated, one with $m_{1}\sim m_{2}\sim 40-57\ \rm M_{\odot}$, and another with $m_{1}\sim 65-85\ \rm M_{\odot}$ and $m_{2}\sim 40-57\ \rm M_{\odot}$. The reason is that Pop~III stars with initial masses $m_{\star}\sim 40-57\ \rm M_{\odot}$ will collapse into BHs directly with no supernovae, while stars with $m_{\star}\sim 85-115\ \rm M_{\odot}$ will also form BHs in the same mass range from PPISNe, such that this mass range is favored, especially for the secondary. We also predict a few percent of NS-BH mergers with $m_{1}\sim 20-85\ \rm M_{\odot}$ and $m_{2}\sim 1.4-3\ \rm M_{\odot}$. Our models predict a broad distribution of $m_{\rm chirp}$ with two peaks. The dominant peak is at $m_{\rm chirp}\sim 50\ \rm M_{\odot}$, and there is a smaller peak at $m_{\rm chirp}\sim 5\ \rm M_{\odot}$ corresponding to NS-BH mergers. The majority of mergers in our models have $m_{\bullet}\sim 20-170\ \rm M_{\odot}$ with a peak in the total mass distribution at $m_{\bullet}\sim 125\ \rm M_{\odot}$.

\begin{figure}
    \centering
    \includegraphics[width=1\columnwidth]{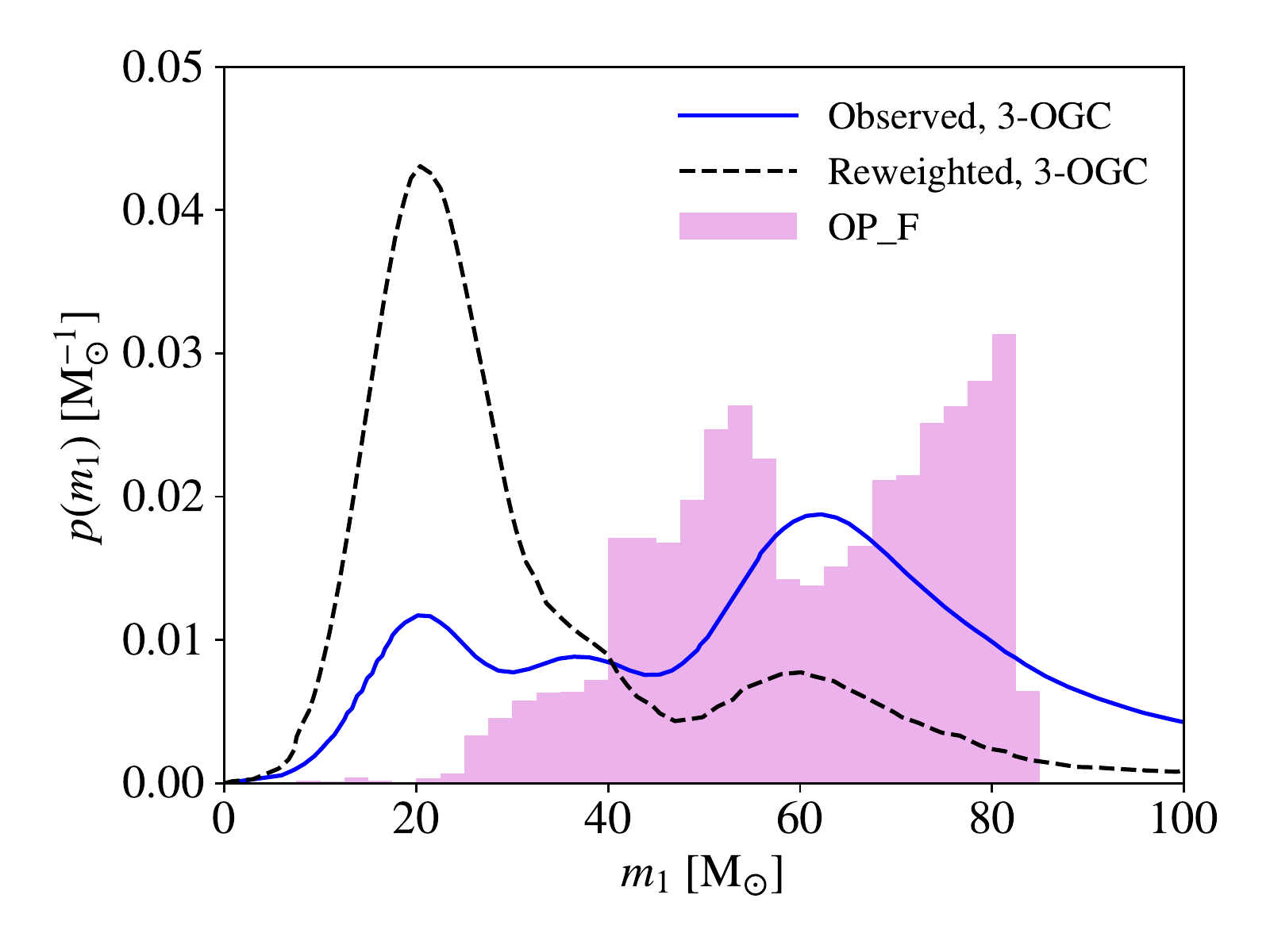}
    \vspace{-25pt}
    \caption{Primary mass distribution of Pop~III binary remnant mergers from the OP\_F model (histograms), compared with the observed (solid) and reweighted (dashed) primary mass distributions inferred from the 3-OGC \citep[see their fig.~5]{3ogc}.}
    \label{m1dis}
\end{figure}

Different from our Pop~III binary remnant mergers, most LIGO events involve relatively less massive BH-BH mergers with $m_{1}\lesssim 50\ \rm M_{\odot}$ and $m_{2}\lesssim 40\ \rm M_{\odot}$. The chirp (total) mass distribution peaks at $\sim 30\ (60)\ \rm M_{\odot}$. This low mass regime is also populated by a few percent of our mergers. Particularly, we have a small peak at $m_{1}\sim 40\ \rm M_{\odot}$, $m_{2}\sim 25\ \rm M_{\odot}$ in the $m_{2}$-$m_{1}$ map. Nevertheless, considering the low local MRD of Pop~III mergers, in terms of overall probability, if there are any LIGO events from Pop~III stars, they are likely to be the most massive ones, such as GW170403, GW190519, GW190602, GW190701, GW190706 and GW190521 (highlighted with triangles in the top panel of Fig.~\ref{mdis}). Actually, the distribution of $m_{1}$ for LIGO events cannot be described by a single truncated power-law, given the presence of features implying a high-mass component at $m_{1}\gtrsim 40\ \rm M_{\odot}$ \citep{abbott2020population,3ogc,roulet2021distribution}. Our results indicate that Pop~III remnants can make important contributions to this high-mass regime, as shown in Fig.~\ref{m1dis}. Moreover, the special event GW190814 with $m_{1}=23_{-2.4}^{+1.6}\ \rm M_{\odot}$ and $m_{2}=2.6_{-0.1}^{+0.2}\ \rm M_{\odot}$, which can be a NS-BH merger, is also covered by our population of Pop~III NS-BH mergers. 

\begin{figure}
    \centering
    \includegraphics[width=1\columnwidth]{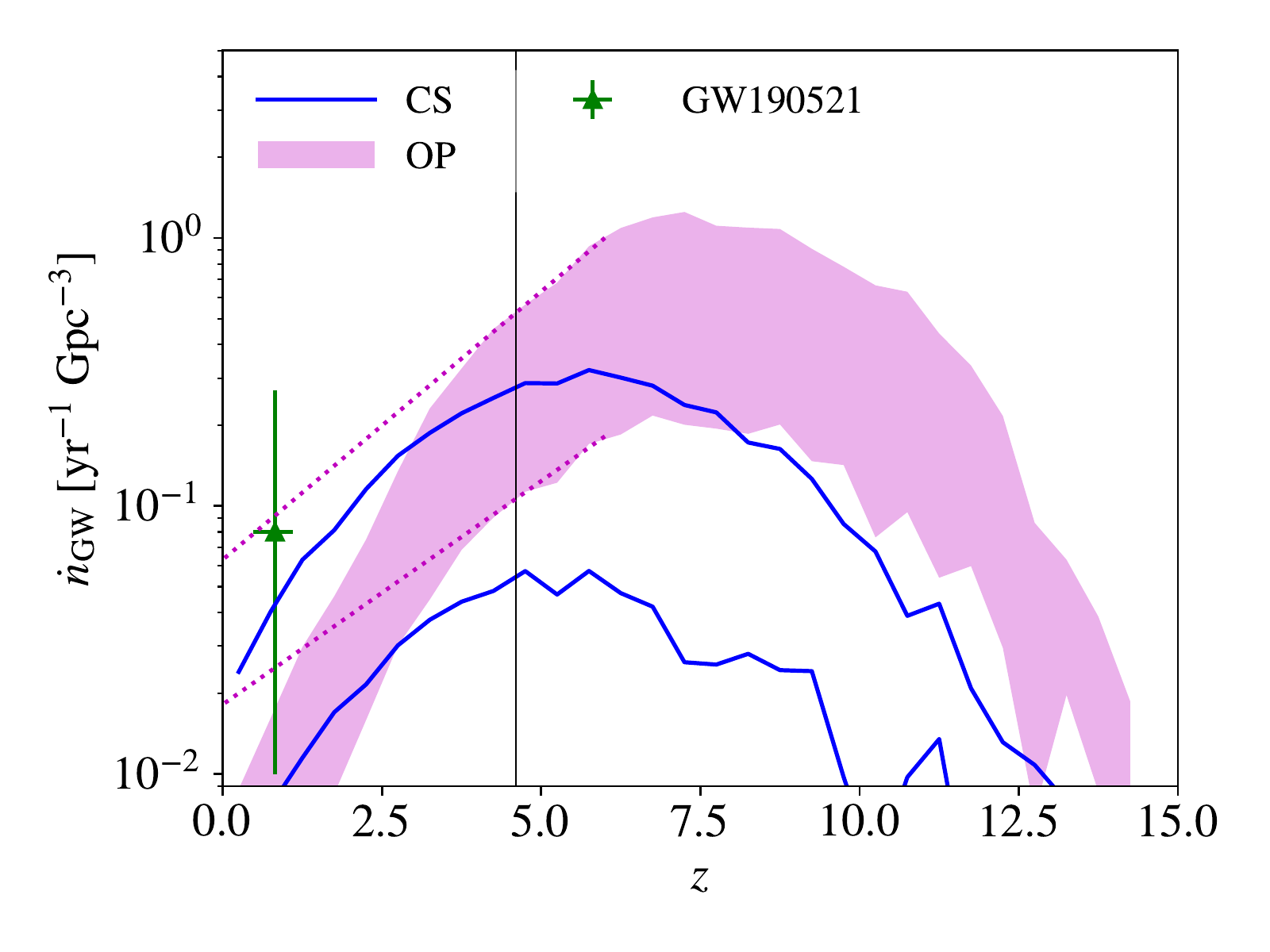}
    \vspace{-25pt}
    \caption{MRDs of GW190521-like mergers from the OP (histograms) and CS models (solid curves), compared with the MRD inferred from GW190521, $\dot{n}_{\rm GW}(z\sim 1)=0.08_{-0.07}^{+0.19}\ \rm yr^{-1}\ Gpc^{-3}$ \citep{lmbh2021}, denoted by the triangle with the 90\% confidence interval. Similar to Fig.~\ref{mrdcomp}, extrapolations of the MRDs from the OP models are shown with the dotted lines. The thin vertical line denotes the turn-over redshift $z\simeq 4.6$ of the target halo below which our results may not be cosmologically representative.}
    \label{gw190521}
\end{figure}

Finally, we focus on the special event GW190521 that involves unusually large BH masses $85_{-14}^{+21}\ \rm M_{\odot}$ and $66_{-18}^{+17}\ \rm M_{\odot}$ \citep{abbott2020gw190521,abbott2020properties,mehta2021,obrien2021detection}, within the standard PPISN mass gap $\sim 55-130\ \rm M_{\odot}$ for Pop~I/II stars (see e.g. \citealt{heger2003massive,belczynski2016effect, woosley2017pulsational,marchant2019pulsational}). This event likely also falls within the high-mass ($m_{i}\gtrsim 50\ \rm M_{\odot}$), low-spin ($\chi_{i}\lesssim 0.2$) regime ($i=1$ or 2) that cannot be easily populated by hierarchical mergers \citep{gerosa2021high}. Several studies have shown that GW190521-like mergers can originate from Pop~III stars \citep{farrell2020gw190521,tanikawa2020population,kinugawa2021formation,bl2020gw190521}, since Pop~III stars can retain most of their hydrogen envelopes (due to their compactness and little mass loss), avoid the PPISN regime, and form BHs in this mass range\footnote{A similar mechanism can also form BHs in the standard mass gap from metal-enriched stars with metallicities up to $0.1\ \rm Z_{\odot}$ with enhanced core overshooting and reduced mass loss \citep{vink2021maximum}.}.

In light of this, we select the events similar to GW190521 from our population of Pop~III mergers in NSCs and calculate their MRDs for different NSC models. Here we define GW190521-like mergers as those with $m_{1}\sim 71-106\ \rm M_{\odot}$ and $m_{\bullet}\sim 133-179\ \rm M_{\odot}$, based on the 90\% confidence intervals from the NRSur7dq4 waveform model \citep{abbott2020gw190521}. The results are shown in Fig.~\ref{gw190521}. We find a MRD range of $\sim 0.01-0.09\ \rm yr^{-1}\ Gpc^{-3}$ at $z\sim 1$, overlapping with that inferred from GW190521, $\dot{n}_{\rm GW}(z\sim 1)=0.08_{-0.07}^{+0.19}\ \rm yr^{-1}\ Gpc^{-3}$ \citep{lmbh2021}, within the 90\% confidence interval. This confirms the conclusion in \citet{bl2020gw190521} that DH in dense star clusters is an important channel of forming massive BH binary mergers like GW190521 with BHs in the PPISN mass gap for Pop~I/II stars.  

\section{Summary and discussions}
\label{s4}
We study the in-fall and evolution of Pop~III binary (compact object) remnants in NSCs in their full cosmological context, based on halo merger trees equipped with self-consistent modelling for SF, stellar feedback, galaxy and NSC properties, DF, hardening by three-body encounters, and GW emission. We find that on average 7.5\% of Pop~III binary remnants will fall into galaxy centres ($<3\ \rm pc$) by DF from field stars. Considering different assumptions on the NSC occupation frequency and hardening efficiency, we then construct the population of Pop~III binary remnant mergers in NSCs and derive their host system statistics, MRDs and mass spectra. We find that $5-50\%$ of the remnants at galaxy centres can merge in NSCs at $z>0$. The dominant ($>90\%$) hosts of these mergers are low-mass ($M_{\rm NSC}\lesssim 10^{6}\ \rm M_{\odot}$) NSCs where the in-fall happens in high-redshift ($z\gtrsim 4.5$) dwarf galaxies with $M_{\star}\lesssim 3\times 10^{7}\ \rm M_{\odot}$. Below we summarize the key features of Pop~III binary remnant mergers formed in this NSC-DH channel and discuss their implications, in comparison with the results for the BSE channel \citep{kinugawa2014possible,hartwig2016,belczynski2017likelihood,hijikawa2021population}, and GW events detected by LIGO \citep{abbott2020population,abbott2020gw190521,3ogc}:

\begin{itemize}
	\item The MRDs from our NSC channel start to rise at $z\sim 13-15$ and peak at lower redshifts $z\sim 5-7$ with comparable peak values $\sim 0.4-10\ \rm yr^{-1}\ Gpc^{-3}$ and local values $\sim 0.02-0.6\ \rm yr^{-1}\ Gpc^{-3}$ compared to the results for the BSE channel, most of which peak at $z\sim 8-13$ and extend to very high redshifts ($z\sim 20$). This difference in time evolution is caused by the delay between initial formation of Pop~III remnants and the emergence of the galaxy/NSC systems required for in-spiral by DF to take place. In this way, the relative importance of the NSC-DH and BSE channels can be evaluated by measuring the MRD at high redshifts ($z\gtrsim 10$) with next-generation instruments. 
        
        \item Unlike the BSE channel, which is only relevant for close binaries ($a_{0}\lesssim 10\ \rm au$), in our NSC-DH channel, the initial binary separations are no longer important such that binaries with very large initial separations (up to $1\rm\ pc$) can still merge in NSCs driven by DH. 
        The NSC-DH channel offers a fundamentally different way of enabling Pop~III mergers even if most Pop~III binaries are initially wide as indicated by N-body simulations of young Pop~III star clusters \citep{liu2021binary}. 
        
        \item Although insensitive to initial binary separations, the NSC-DH channel is sensitive to the remnant masses and thus the Pop~III IMF. Note that both DF and GW emission are more efficient for more massive binaries such that the fraction of remnants merged at $z>0$ in NSCs decreases from $\sim 7\%$ for $m_{\bullet}\sim 100\ \rm M_{\odot}$ to $\sim 0.2\%$ for $m_{\bullet}\sim 10\ \rm M_{\odot}$, and the MRD of Pop~III binary remnant mergers in NSCs is lower when the IMF is less top-heavy. For instance, in this work, we consider a log-flat Pop~III IMF with an upper bound of $170\ \rm M_{\odot}$.  Given a fixed total stellar mass, if Pop~III stars were less massive with an upper mass limit of $20\ \rm M_{\odot}$, the MRD in the NSC-DH channel would be reduced by a factor of 15, while that of the BSE channel would be increased by a factor of 5. 
        
        \item Lacking close binary interactions, Pop~III binary remnant mergers formed in the NSC-DH channel will have different binary properties with respect to those formed via BSE. 
        \begin{itemize}
            \item First, we expect more binaries with small mass ratios without the regulation of binary mass transfer.
            \item {Second, the NSC-DH channel can produce eccentric mergers efficiently since (initially) wide remnant binaries will not be pre-circularized by BSE, and three-body encounters enhance eccentricity. For instance, under $\kappa=0.1$ and $\Delta_{\rm c}=100$, $\sim 0.1\%$ of our Pop~III remnant binaries with high initial eccentricities ($e_{0}\gtrsim 0.999$) have $e\gtrsim 0.1 (f_{\rm GW}/0.1\ \rm Hz)^{-1}$ when observed at a rest-frame frequency $f_{\rm GW}$, given typical binary parameters $m_{1}\sim 70\ \rm M_{\odot}$, $m_{2}\sim 40\ \rm M_{\odot}$ and $a_{0}\sim 300\ \rm au$ in NSCs of $M_{\rm NSC}\sim 10^{6}\ \rm M_{\odot}$, $R_{\rm NSC}\sim 3\ \rm pc$ and $\gamma_{\rm NSC}\sim 1.5$. Such eccentric sources can be identified with $e\gtrsim 10^{-4}-0.1$ at $z\lesssim 10$ by 3rd generation GW detectors that are most sensitive at $f_{\rm GW}\sim 0.1-10\ \rm Hz$.}
            \item Finally, the spin distributions can also be different. In this work we do not consider the spins of remnants for simplicity. Here we briefly discuss the key processes that determine the spin distributions for the two channels. 
        It is found in small-scale hydrodynamic simulations (e.g. \citealt{stacy2010first,stacy2012first,stacy2016building,greif2012formation,
        stacy2013constraining,susa2014mass,machida2015accretion,hirano2017formation,
        susa2019merge,sugimura2020birth,chiaki2020}) that Pop~III stars form in small groups from disk fragmentation. In this case, the initial spins of individual stars and binary orbit vectors will be mostly aligned imperfectly as the angular momenta are inherited from the same disk and meanwhile affected by local flows during fragmentation and mergers as well as N-body dynamics. In the NSC-BH channel, the binary orbital plane can be tilted by dynamical interactions (see e.g. \citealt{trani2021spin}), while for close binaries in the BSE channel, the stellar spins are regulated by tidal effects and mass transfer, whose effects are rather complex (see e.g. \citealt{kinugawa2020,tanikawa2021merger,stegmann2021flipping}). Natal kicks can also alter the binary orbit vector with respect to the spins \citep{tanikawa2021merger}. 
        \end{itemize}
        { We defer a detailed analysis of the detectable eccentricities and spins of Pop~III binary remnant mergers for the NSC-DH channel to future work.}
        
        \item Compared with LIGO events mostly involving BHs with $m_{\rm rem}\lesssim 50\ \rm M_{\odot}$, our Pop~III mergers are generally more massive with $m_{\rm rem}\sim 40-85\ \rm M_{\odot}$. Considering the local MRD inferred from LIGO events, $\dot{n}_{\rm GW}(z\sim 0.33)=23.9_{-8.6}^{+14.3}\ \rm yr^{-1}\ Gpc^{-3}$ \citep{abbott2020population}, Pop~III remnants mergers from our NSC-DH channel can count for up to a few percent of the 57 events detected so far \citep{3ogc}, particularly the most massive ones. In general we expect the low-redshift ($z\lesssim 2$) regime to be dominated by mergers from Pop~I//II progenitors, while detection of Pop~III mergers with be most promising at high redshifts ($z\gtrsim 5$). Actually, most of our Pop~III binary remnant mergers in NSCs can be detected by the 3rd generation of future GW instruments, capable of reaching $z\gtrsim 10$, such as the Einstein Telescope \citep{hild2009xylophone} and Decihertz Observatory \citep{dechihertz}, for which we predict a full-sky detection rate of $\gtrsim 170-2720\ \rm yr^{-1}$. 
        \item Our models can produce BH-BH mergers similar to the special event GW190521 \citep{abbott2020gw190521,abbott2020properties,mehta2021,obrien2021detection} that involves two BHs in the standard PPISN mass gap $\sim 55-130\ \rm M_{\odot}$ for Pop~I/II stars (see e.g. \citealt{heger2003massive,belczynski2016effect, woosley2017pulsational,marchant2019pulsational}). The predicted MRD for such mergers agrees with that inferred from GW190521 within the 90\% confidence interval. This further supports the proposal that mergers of BHs in the standard PPISN mass gap can originate from Pop~III stars \citep{farrell2020gw190521,tanikawa2020population,kinugawa2021formation,bl2020gw190521}. Moreover, the NSC-DH channel does not suffer from the possible reduction of BH mass due to close binary interactions in the BSE channel \citep{farrell2020gw190521}. Actually, at low redshifts when mergers from different formation channels are mixed together, the mass gaps for (P)PISNe can be useful windows to distinguish the mergers from different stellar populations since the location of mass gap depends on metallicity\footnote{The exact dependence of mass gap location on metallicity is still uncertain in current stellar evolution models due to uncertainties in e.g. mass loss and convective overshooting.} \citep{spera2017very,heger2010nucleosynthesis,abbott2020gw190521,abbott2020properties,vink2021maximum}. This is particularly relevant for mergers involving high-mass ($\gtrsim 50\ \rm M_{\odot}$), low-spin ($\lesssim 0.2$) BHs that are not likely to be hierarchical mergers \citep{gerosa2021high}. In our case, Pop~III stars can be the unique progenitors for mergers with low-spin BHs in the mass range of $\sim 55-85\ \rm M_{\odot}$. {For mergers involving at least one BH in this mass range, our models predict a full-sky detection rate of $\sim 110-1640\ \rm yr^{-1}$ for $z_{\max}=10$. }
\end{itemize}

In general, our results show that the NSC-DH channel is efficient at producing Pop~III binary remnant mergers, which not only complements the well-known BSE channel, but also highlights the importance of environment for the evolution of compact object binaries. The GW events from this channel will be detected at promising rates in the next decades. Once distinguished from other populations by their unique features with a large enough sample, they will become a useful probe to the first stars and also high-$z$ NSCs formed in the first galaxies. In concluding, we would like to discuss the caveats of our model and suggest avenues for future work:

\begin{itemize}
    \item Our merger trees are targeted towards a MW-like halo virialized at $z\simeq 2.5$, such that they are only cosmologically representative at high redshifts ($z\gtrsim 5$, before the turn-over of the target halo). As a result, the population of Pop~III binary remnant mergers in NSCs at $z\lesssim 5$ may be incomplete in our models. And we have used simple assumptions and extrapolation to estimate the MRD in this low-redshift regime. In future work, we will generate a broader range of target haloes at $z=0$ and combine their results weighted by the halo mass function to construct a cosmologically representative population of Pop~III binary remnant mergers in NSCs down to $z=0$ (see e.g. \citealt{magg2016new}).
    \item As a common shortcoming of merger trees generated by the extended Press-Schechter formalism \citep{parkinson2008generating,mo2010galaxy}, positional/clustering information of haloes is lacking in our stellar feedback model, such that radiative feedback can only be described by a uniform background, and metal enrichment is implemented in a stochastic manner. { Actually, the stochastic metal enrichment model in HT15 does not achieve numerical convergence with redshift/time resolution such that we have further introduced a scaling factor for the metal-enriched volume to calibrate our predicted SFRDs with those in simulations and observations.} Note that stellar feedback regulates not only the overall Pop~III SF history, but also the host haloes and basic units of Pop~III SF that are important for binary statistics \citep{liu2020did,liu2021binary}. { Future studies should adopt more advanced feedback models taking into account local effects with merger trees constructed from cosmological simulations in which the positions of haloes are available (see e.g. \citealt{magg2018predicting,hartwig2018descendants}).}
    \item Crucially, as DF and DH are complex processes in the context of galaxy formation and evolution, our semi-analytical approach relies on idealized assumptions for galaxy and NSC properties based on empirical constraints in the local Universe. In this way, we are only able to predict a broad MRD range for the Pop~III NSC-DH channel by considering limits for idealized (optimistic and pessimistic) scenarios. The detailed complex physics involved in the in-fall and evolution of compact object binaries in NSCs, such as galaxy dynamics and mergers, NSC formation, growth and structure, as well as central massive BHs, can only be modelled self-consistently with cosmological hydrodynamic simulations. Simulating structure formation with these physical processes included and enough resolution for Pop~III SF in a representative cosmological volume (down to $z=0$) is a challenging but promising task. As an exploratory step in this direction, we plan to bulid sub-grid models for NSCs and central massive BHs based on existing analytical models (e.g. \citealt{devecchi2009formation,devecchi2010high,devecchi2012high}), and sub-grid models designed for cosmological simulations with limited resolution (e.g. the E-MOSAICS project, \citealt{pfeffer2018mosaics}). These models will be implemented and tested in meso-scale simulations (in a co-moving volume of a few hundred $\rm Mpc^{3}$) that can resolve minihaloes \citep[][]{jaacks2019}, coupled with existing schemes for Pop~III and Pop~II SF and stellar feedback (see e.g. \citealt{johnson2013first,xu2016late,sarmento2018following,liu2020gw,liu2020did}).
\end{itemize}

Identifying observational probes of the elusive first generation of stars has been a long-term challenge, given that any such signatures are typically hidden in the dominant foreground from subsequent SF. Actually, Pop~III stars, formed mostly before reionization ($z\gtrsim 6$), only make up $\sim 10^{-5}$ of all stars (ever formed) in the Universe. However, as shown by our results for the NSC-DH channel and previous studies for the BSE channel, a much higher fraction ($\sim 10^{-3}-0.1$) of massive ($m_{\bullet}\gtrsim 20\ \rm M_{\odot}$) compact object mergers can originate from Pop~III stars, including those observed by LIGO in the local Universe. This is a natural consequence/extrapolation from the strong metallicity dependence of the efficiency of producing massive compact object binaries found in Pop~I/II stars (e.g. \citealt{dominik2015,giacobbo2018progenitors,mapelli2018host,mapelli2019properties}). In the next decades when 3rd-generation GW detectors map the demography of BHs and NSs with thousands of compact object mergers, `gravitational-wave archaeology', targeting the massive BH remnants from Pop~III, will provide an unusually efficient channel to test theories of the first stars.

\section*{Acknowledgements}
The authors would like to thank Tilman Hartwig for helpful comments. This work was supported by National Science Foundation (NSF) grant AST-1413501. 


\section*{Data availability}
The data underlying this article will be shared on reasonable request to the corresponding authors.

\bibliographystyle{mnras}
\bibliography{ref} 

\appendix


\label{lastpage}
\end{document}